\documentclass[12pt]{emulateapj} 
\usepackage{amsmath}
\usepackage{amssymb}
\usepackage{amstext}
\usepackage{graphicx}
\usepackage{epstopdf}
\usepackage{epsfig}
\usepackage{color}
\definecolor{myColor}{rgb}{0.9,0.9,0.9}    
\begin{document}
\renewcommand\bottomfraction{.9}
\shorttitle{Exoplanet atmosphere retrieval}
\title{A Temperature and Abundance Retrieval Method for Exoplanet Atmospheres}
\author{N. Madhusudhan\altaffilmark{1} \& S. Seager\altaffilmark{1}$^{,}$\altaffilmark{2}}
\altaffiltext{1}{Department of Physics and Kavli Institute for 
Astrophysics and Space Research, MIT, Cambridge, MA 02139; {\tt 
nmadhu@mit.edu}
} 
\altaffiltext{2}{Department of Earth, Atmospheric, and Planetary
Sciences, MIT, Cambridge, MA 02139;} 

\begin{abstract}

We present a new method to retrieve molecular abundances and
temperature profiles from exoplanet atmosphere photometry and spectroscopy. We
run millions of 1D atmosphere models in order to cover the large range
of allowed parameter space.  In order to run such a large number of 
models, we have developed a parametric pressure-temperature ($P$-$T$) 
profile coupled with line-by-line radiative transfer, hydrostatic
equilibrium, and energy balance, along with prescriptions for non-equilibrium 
molecular composition and energy redistribution. The 
major difference from traditional 1D radiative transfer models is the
parametric $P$-$T$ profile, which essentially means adopting energy 
balance only at the top of the atmosphere and not in each layer. 
We see the parametric $P$-$T$ model as a parallel approach to the traditional 
exoplanet atmosphere models that rely on several free parameters to encompass 
unknown absorbers and energy redistribution. The parametric 
$P$-$T$ profile captures the basic physical features of temperature structures 
in planetary atmospheres (including temperature inversions), and fits a 
wide range of published $P$-$T$ profiles, including those of solar system planets. 
We apply our temperature and abundance retrieval method to 
the atmospheres of two transiting exoplanets, HD~189733b and HD~209458b, which 
have the best {\it Spitzer} and {\it HST} data available. For HD~189733b, we find 
efficient day-night redistribution of energy in 
the atmosphere, and molecular abundance constraints confirming the 
presence of H$_2$O, CO, CH$_4$ and CO$_2$. For HD~209458b, 
we confirm and constrain the day-side thermal inversion in an 
average 1D temperature profile. We also report independent detections of 
H$_2$O, CO, CH$_4$ and CO$_2$ on the dayside of HD~209458b, based on 
six-channel {\it Spitzer} photometry. We report constraints for HD~189733b 
due to individual data sets separately; a few key observations are variable in 
different data sets at similar wavelengths. Moreover, a noticeably 
strong CO$_2$ absorption in one data set is significantly weaker in another. We 
must, therefore, acknowledge the strong possibility that the atmosphere is variable, 
both in its energy redistribution state and in the chemical abundances.
\end{abstract}


\section{Introduction}

Major advances in the detection and characterization of extrasolar
planet atmospheres have been made during the last decade. Over a dozen
hot Jupiter atmospheres have been observed by the {\it Spitzer Space Telescope} 
and a handful by the {\it Hubble Space Telescope} ({\it HST}). These
observations have changed the field of exoplanets, enabling for the
first time studies of atmospheres of distant worlds.

Observational highlights include the identification of molecules
and atoms, and signatures of thermal inversions. {\it HST} observed 
water vapor and methane in transmission in HD~189733b (Swain et al. 2008b).  
Water vapor was also inferred in transmission at the 3.6 $\micron$, 
5.8 $\micron$, and 8 $\micron$ in HD~189733b (Tinetti et al. 2007) 
(but c.f. Desert et al. 2009, and also Beaulieu et al. 2009). Water was inferred on the dayside 
of HD~189733b from the planet-star flux contrast in the 
3.6 $\micron$ - 8.0 $\micron$ {\it Spitzer} broadband photometry 
(Barman 2008; Charbonneau et al. 2008). Additionally, Grillmair et al (2008) 
reported a high signal-to-noise spectrum of the dayside of HD~189733b 
in the 5 $\micron$ - 14 $\micron$ range, using the {\it Spitzer} Infrared 
Spectrograph (IRS), and reported detection of water. More recently, {\it HST} 
detected carbon dioxide in thermal emission in HD~189733b (Swain et al. 2009a), 
in 1.4 $\micron$ - 2.6 $\micron$. Previously, sodium was detected in 
HD~209458b (Charbonneau et al. 2002) and HD~189733b (Redfield et
al. 2008). Several observations of the planetary dayside have also  been 
reported for HD~209458b, first observed by {\it Spitzer} at 24 
$\micron$ (Deming et al. 2005 and Seager et al. 2005). A major observational 
discovery was the finding of strong emission features in the broadband photometry of 
HD~209458b at secondary eclipse (Knutson et al. 2008), implying a thermal inversion in the atmosphere (Burrows et al. 2007 \& 2008). In tandem with our present work, water, methane 
and carbon dioxide were independently inferred and recently reported on the dayside 
of HD~209458b, with {\it HST} NICMOS and five-channel {\it Spitzer} photometry (Swain et al. 2009b).
Additionally, there are hints that variability of hot Jupiter thermal emission may be common 
(Knutson et al. 2007 vs. Charbonneau et al. 2008; Charbonneau et al. 2008 vs. Grillmair et al. 2008;  
Deming et al. 2005 vs. Deming, private communication 2009). The discoveries of 
atmospheric constituents and temperature structures mark the remarkable 
successes of exoplanet atmosphere models in interpreting the observations. 

Despite the successes of model interpretation of data, major 
limitations of traditional self-consistent atmosphere models are 
becoming more and more apparent. Model successes include the 
detection of thermal inversions on the day-side of hot Jupiters (Burrows et al. 2007;
Knutson et al. 2008), and subsequent classification of systems on the
same basis (Burrows et al. 2008; Fortney et al. 2008; Seager et al. 2008). 
The nature of the absorbers causing the inversions, however, is not 
yet known (Knutson et al. 2008 \& 2009a), forcing modelers to use an adhoc 
opacity source to explain the data. The intially favored opacity sources, TiO 
and VO may be unlikely to cause inversions of the observed magnitude 
(Spiegel et al. 2009). 

A second model limitation arises because 1D radiative-convective 
equilibrium models cannot include the complex physics involved in 
hydrodynamic flows. Existing models use a 
parameterization of energy transfer from the day side to night side
(e.g., Burrows et al. 2008), using parameters for the locations of energy 
sources and sinks, and for the amount of energy transported. 

One further example of atmospheric model limitations involves the 
treatment of chemical compositions. Self-consistent models generally 
calculate molecular mixing ratios based on chemical equilibrium along 
with the assumption of solar abundances, or variants thereof (e.g., Seager et
al. 2005). However, hot Jupiter atmospheres should host manifestly 
non-equilibrium chemistry (e.g., Liang et al. 2003, Cooper \& Showman, 2006), 
which render the assumption of chemical equilibrium only a fiducial starting point.

With parameters used to cover complicated or unknown physics or 
chemistry, existing ``self-consistent'' radiative-convective equilibrium models
are no longer fully self consistent. Furthermore, given the computational 
time of such models, only a few models are typically published to 
interpret the data, often only barely fitting the data ---a quantitative 
measure of fit is typically absent in the literature.  In an ideal world, one 
would construct fully self-consistent 3D atmosphere models (e.g., 
Showman et al. 2009) that include hydrodynamic flow, radiative transfer, 
cloud microphysics, photochemistry and
non-equilibrium chemistry, and run such models over all of parameter
space anticipated to be valid. Such models will remain idealizations
until computer power improves tremendously, given that atmospheric
circulation models can take weeks to months to run even with simple 
radiative transfer schemes.

We are motivated to develop a data-interpretation framework for exoplanet 
atmospheres that enables us to run millions of models in order to 
constrain the full range of pressure-temperature ($P$-$T$) profiles
and abundance ranges allowed by a given data set. In this work, we report 
a new approach to 1D modeling of exoplanet atmospheres. Taking a cue from
parameterized physics already being adopted by existing models, we go much
further by parameterizing the $P$-$T$ profile as guided by basic physical 
considerations and by properties of $P$-$T$ profiles of planetary atmospheres 
in the literature. Indeed, it appears at present, the only way to be able to run enough 
models to constrain the $P$-$T$ structure and abundances is to use a parameterized
$P$-$T$ profile. The essential difference between our new method and
currently accepted radiative-convective equilibrium models is in the
treatment of energy balance. We ensure global energy balance at the
top of the atmosphere, whereas conventional atmosphere models use
layer-by-layer energy balance by way of radiative and convective
equilibrium. Our new method can be used as a stand-alone model, 
or it can be used to identify the parameter space in which to run a reasonable 
number of traditional model atmospheres. We call our method 
``temperature and abundance retrieval" of exoplanet atmospheres. 

Atmospheric temperature and abundance retrieval is not new (e.g. Goody \&
Yung, 1989). Studies of planetary atmospheres in the solar system 
use temperature retrieval methods, but in the present context there is 
one major difference. Exquisite data for solar system planet atmospheres means a
fiducial pressure - temperature profile can be derived. The temperature 
retrieval process for atmospheres of solar system planets therefore 
involves perturbing the fiducial temperature profile. 
For exoplanets, where data are inadequate, there is no starting point to derive 
a fiducial model. Nevertheless, Swain et al. (2008b, 2009a) used previously published 
model $P$-$T$ profiles of HD~189733b, and varied molecular abundances 
to report model fits to {\it HST} NICMOS spectra of HD~189733b. In another 
study, Sing et al. (2008) used a parametric temperature profile with a thermal 
inversion, and parametrized abundances, required specifically to explain 
{\it HST} STIS optical observations of HD~209458b. Our model, however, is completely 
general in the choice of $P$-$T$ profiles, ranges over tens of thousands 
of profiles, and satisfies the physical constraints of global energy balance 
and hydrostatic equilibrium. Our model reports, with contour 
plots, the quantitatively allowed ranges of $P$-$T$ profiles and molecular 
abundances. In addition, we also report constraints on the albedo and 
day-night energy redistribution, and on the effective temperature.

In this paper, the $P$-$T$ profiles are introduced in \S~2, and the
radiative transfer model and parameter space search strategy are
described in \S~3. Results of data interpretation with the model are
presented in \S~4 for HD~189733b, and in \S~5 for HD~209458b. \S~6
discusses the overall approach, and prospects for future work. 
And, in \S~7 we present a summary of our results and conclusions. 

\section{Temperature - Pressure Structure}

Our goal is to introduce a new method of parametrizing one-dimensional
(1-D) models of exoplanetary atmospheres. Our approach is to
parametrize the pressure-temperature ($P$-$T$) profiles, instead of
separately parametrizing each of the different physical processes that
contribute to the $P$-$T$ profiles. In what follows, we will first
review the basic physics behind the temperature structure of planetary
atmospheres which serves as motivation for our $P$-$T$ form.  Next, we
will describe our parametric $P$-$T$ profile. Finally, we compare our
model $P$-$T$ profiles with published profiles for some known
planetary systems.

\subsection{Temperature Structure in Planetary Atmospheres}

The pressure-temperature profiles of planetary atmospheres are 
governed by basic physics. Here, we qualitatively describe the physics and 
shape of a representative $P$-$T$ profile starting from the bottom of the 
atmosphere. We focus on the generality that the temperature structure at 
a given altitude depends on the opacity at that altitude, along with density 
and gravity.

In the deepest layers of the planet atmosphere, convection is the
dominant energy transport mechanism. The high pressure (equivalently,
high density) implies a high opacity, making energy transport by 
convection a more efficient energy transport mechanism than radiation.
For hot Jupiters, the dayside radiative - convective boundaries usually occur at
the very bottom layers of the atmospheres, at pressures higher than
$\sim$ 100 bar.  In the layers immediately above this boundary, the optical
depth is still high enough that the diffusion approximation of
radiative transport holds. The diffusion approximation, and the 
incident stellar flux predominantly absorbed higher up in the atmosphere, 
lead to a nearly isothermal temperature structure in the layers immediately 
above the radiative-convective boundary in hot Jupiters. It is assumed here 
that the energy sources in the planet interior are weak compared to the stellar irradiation.
In the solar system giant planet atmospheres (and cooler exoplanet atmospheres 
as yet to be observed), the radiative - convective boundary occurs at lower
pressures in the planet atmosphere than for hot Jupiters.

Above the isothermal diffusion layer, i.e. at lower pressures, 
the atmosphere becomes optically thin and the diffusion approximation breaks
down. These optically thin layers are at altitudes where thermal inversions can be formed, 
depending on the level of irradiation from the parent star and the presence of 
strong absorbing gases or solid particles. Thermal inversions, or ``stratospheres'',
are common to all solar system planets and have recently been
determined to exist in several hot Jupiter atmospheres 
(Burrows et al. 2008). For the solar system giant planets, 
photochemical haze due to CH$_4$ is the primary 
cause of stratospheres, and for Earth it is O$_3$. For hot 
Jupiters, TiO and VO may be responsible for the thermal inversion (but see 
Spiegel et al. 2009), but the identification of the absorbers is still 
under debate. At still lower pressures, below P $\sim10^{-5}$ bar, 
the optical depths eventually become so low that the layers of 
the atmosphere are transparent to the incoming and outgoing 
radiation. 

\subsection{A Parametric $P$-$T$ Profile}
\label{sec-parametricPT}
We propose a parametric $P$-$T$ profile, motivated by physical
principles, solar system planet $P$-$T$ profiles, and 1D
``self-consistent'' exoplanet $P$-$T$ profiles generated from model
atmosphere calculations reported in the literature. Our ``synthetic"
$P$-$T$ profile encapsulates stratospheres, along with the low and high
pressure regimes of the atmosphere. We will refer
to our synthetic $P$-$T$ profile as the ``parametric $P$-$T$ profile".

\begin{figure}[ht]
\begin{center}
\includegraphics[width=0.5\textwidth]{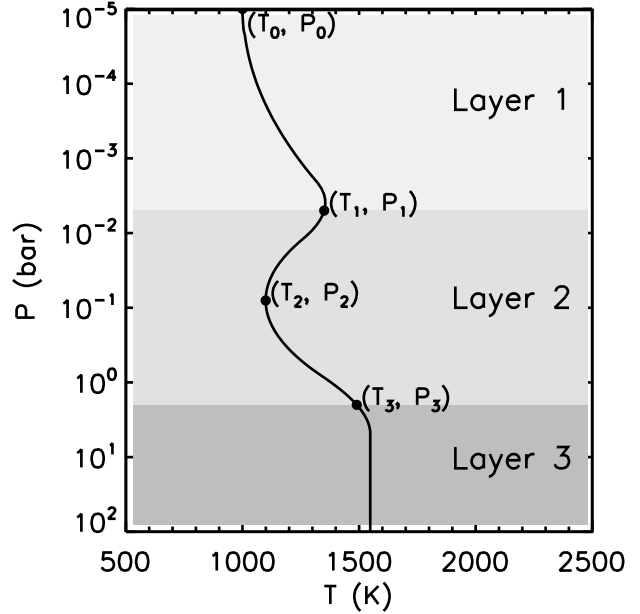}
\caption{The parametric pressure-temperature ($P$-$T$) profile. In 
the general form, the profile includes a thermal inversion layer
(Layer 2) and has six free parameters, see (\ref{eqn:pt_profile}) . 
For a profile with thermal inversion, $P_2 > P_1$, and for one 
without a thermal inversion, $P_1 \ge P_2$. An isothermal profile is 
assumed for pressures above P$_3$ (Layer 3). Alternatively, 
for cooler atmospheres with no isothermal layer, Layer 2 could 
extend to deeper layers and Layer 3 could be absent (see \S~\ref{sec-fit_pt}). }
\label{fig:pt_schematic}
\end{center}
\end{figure}

\begin{figure*}[ht]
\centering
\includegraphics[width = \textwidth]{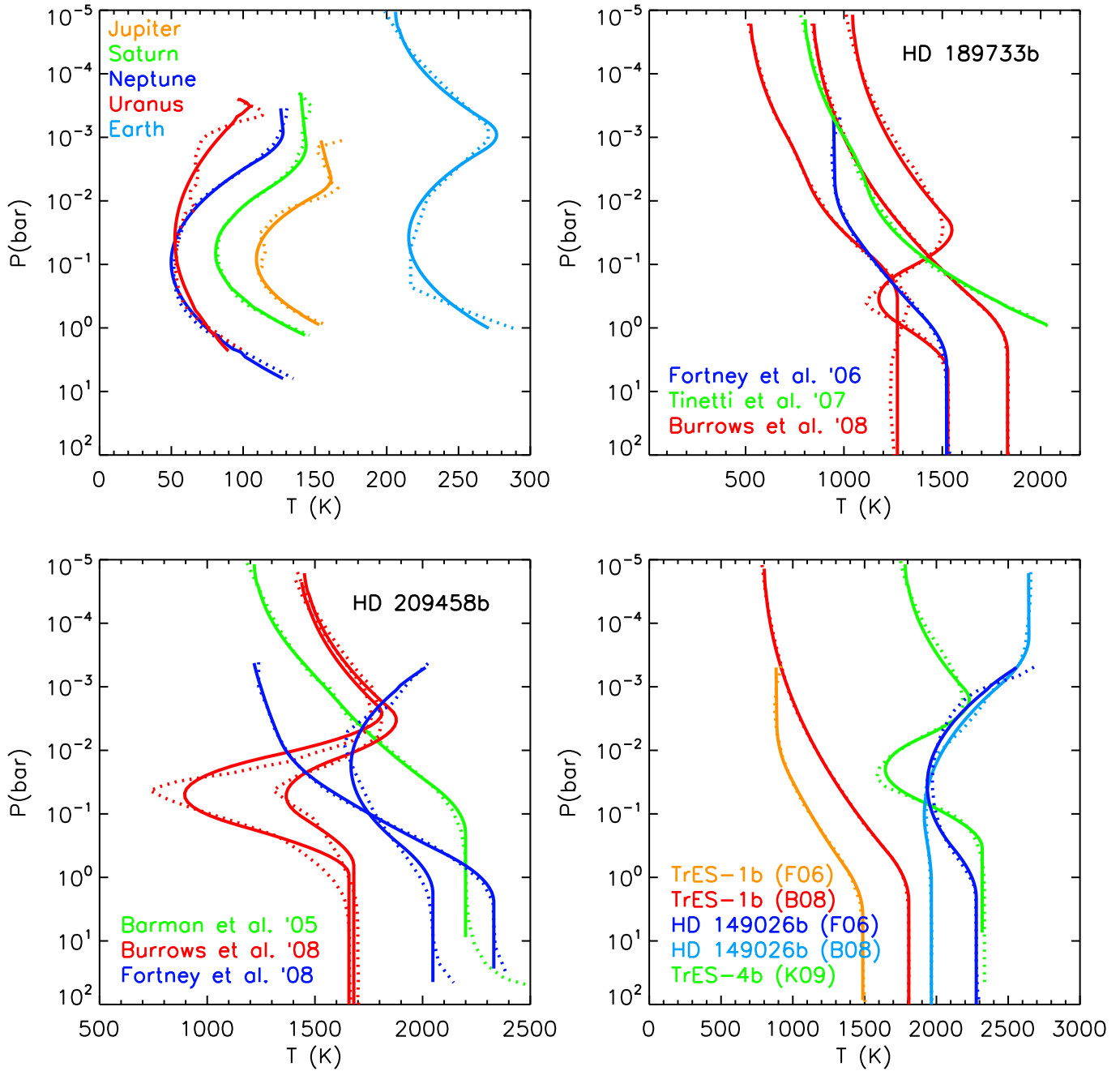}
\caption{Comparison of the parametric $P$-$T$ profile with
previously published profiles. In each case, the dotted line is the
published $P$-$T$ profile and the solid line is a fit with the 
parametric profile (see \S~\ref{sec-fit_pt}). The $P$-$T$ profiles 
of solar-system planets (upper-left panel) were obtained from the 
NASA Planetary Data System (see text for references). 
In the lower-right panel, ``B08", ``F06" and ``K09" refer to 
Burrows et al. (2008), Fortney et al. (2006) and Knutson et al. 
(2009a), respectively.}
\label{fig:fit_profs}
\end{figure*}

The atmospheric altitudes of interest are those where the spectrum is
formed.  Nominally, we consider this range to be between $10^{-5}$ 
bar $\leq P \leq 100$ bar, and we will refer to this pressure range as
the ``atmosphere" in the rest of the paper. At pressures lower than 
$10^{-5}$ bar, the atmosphere layers are nearly transparent to the
incoming and outgoing radiation at visible and infrared
wavelengths. And, above pressures of $\sim100$ bar (or even
above pressures of $\sim 10$ bar), the optical depth is too high for any
radiation to escape without being reprocessed. The corresponding 
layers do not contribute significantly to the emergent spectrum. Figure~\ref{fig:pt_schematic} shows a schematic parametric $P$-$T$ profile. In our model, 
we divide the atmosphere into three representative zones or ``Layers" 
as shown in Figure~\ref{fig:pt_schematic}. The upper-most layer, Layer 1, in our model profile is a ``mesosphere'' with no thermal inversions. The middle layer, Layer 2, 
represents the region where a thermal inversion (a ``stratosphere'') is possible. 
And, the bottom-most layer, Layer 3, is the regime where a high 
optical depth leads to an isothermal temperature structure. Layer 3 
is used with hot Jupiters in mind; for cooler atmospheres this layer 
can be absent, with Layer 2 extending to deeper layers. 

Our proposed model for the $P$-$T$ structure in Layers 1 and 2 is a
generalized exponential profile of the form:
\begin{equation} 
  P = P_oe^{\alpha(T - T_o)^\beta}, 
  \label{eqn:pt_structure}
\end{equation}
where, $P$ is the pressure in bars, $T$ is the temperature in K, and
$P_o$, $T_o$, $\alpha$ and $\beta$ are free parameters.
For Layer 3, the model profile is given by $T = T_3$, where $T_3$ 
is a free parameter. 

Thus, our parametric $P$-$T$ profile is given by:
\begin{align}
  P_0 &< P < P_1&:\hspace{0.7cm} &P = P_0e^{\alpha_1(T - T_0)^{\beta_1}} &\hspace{0.2cm} \rm{Layer 1} \nonumber \\
  P_1 &< P < P_3&:\hspace{0.7cm} &P = P_2e^{\alpha_2(T - T_2)^{\beta_2}} &\hspace{0.2cm} \rm{Layer 2}  \label{eqn:pt_profile} \\
  P~&> P_3&:\hspace{0.7cm} &T = T_3 &\hspace{0.2cm} \rm{Layer 3}  \nonumber
\end{align}

In this work, we empirically set $\beta_1 = \beta_2 = 0.5$ 
(see \S \ref{sec-fit_pt}). The space of $P$-$T$ profiles is spanned in all 
generality with the remaining parameters, rendering $\beta$ redundant 
as a free parameter. Then, the model profile in (\ref{eqn:pt_profile}) 
has nine parameters, namely, $P_0$, $T_0$, $\alpha_1$, $P_1$, 
$P_2$, $T_2$, $\alpha_2$, $P_3$, and $T_3$. Two of the parameters 
can be eliminated based on the two constraints of continuity at the 
two layer boundaries, i.e., Layers 1--2 and Layers 2--3. And, in 
the present work, we set $P_0 = 10^{-5}$ bar, i.e., at the top of our 
model atmosphere. Thus, our parametric profile in its complete 
generality has six free parameters. 

Our $P$-$T$ profile consists of 100 layers in the pressure range 
of $10^{-5} - 100$ bar, encompassing the three zonal ``Layers" 
described above. The 100 layers are uniformly spaced in $\log(P)$. 
For a given pressure, the temperature in that layer is determined from equation 
(\ref{eqn:pt_profile}), using the form $T = T(P)$. The kinks at 
the layer boundaries are removed by averaging the profile with 
a box-car of 10 layers in width. 

\subsection{Comparison with known $P$-$T$ Profiles}
\label{sec-fit_pt}
The parametric $P$-$T$ profile described in \S~\ref{sec-parametricPT}
fits the actual temperature structures of a wide variety of 
planetary atmospheres. Figure~\ref{fig:fit_profs} shows the comparison of our model $P$-$T$
profile with published $P$-$T$ profiles of several solar system planets
and hot Jupiters. For each case, the published profile was fitted
with our six-parameter $P$-$T$ profile using a Levenberg-Marquardt fitting
procedure (Levenberg, 1944; Marquardt, 1963). 
We were able to find best fits to all the published profiles with the
$\beta$ parameter fixed at 0.5; indicating that $\beta$ is a redundant
parameter. Therefore, we set $\beta_1 = \beta_2 = 0.5$ for all the $P$-$T$ 
profiles in this work. In addition, it must be noted that, for the parametric $P$-$T$ profile 
to have a thermal inversion, the condition $P_2 > P_1$ must be satisfied, and 
for one without a thermal inversion, $P_1 \ge P_2$.

For the solar system planets (top left panel of Figure~\ref{fig:fit_profs}), published profiles
show detailed temperature structures obtained via several direct 
measurements and high-resolution temperature retrieval methods 
(Lindal et al. 1981; Lindal et al. 1985; Lindal et al. 1987; Lindal et al. 1990; 
U.S. Standard Atmosphere 1976; ${\rm http://atmos.nmsu.edu/planetary\_datasets/}$). 
For hot Jupiters, on the other hand, observations are limited. Consequently, the published 
profiles for hot Jupiters (Figure~\ref{fig:fit_profs}) are obtained from self-consistent 
1-D models reported in the literature attempting to explain the available 
data on each system. As is evident from Figure~\ref{fig:fit_profs}, the parametric 
$P$-$T$ profile provides good fits to all the known profiles. 
We emphasize that, while the fits alone are not a test of our retrieval 
method, they show that our parametric $P$-$T$ profile is general enough to fit the 
wide range of planetary atmospheres represented in Figure~\ref{fig:fit_profs}. 

We also tested the emergent spectrum obtained with our parametric 
$P$-$T$ profile by comparing it with a self-consistent model. In this 
regard, we used a $P$-$T$ profile from a self-consistent atmosphere 
model (Seager et al. 2005) to generate an emergent spectrum for HD~209458b. 
The emergent spectrum calculated by our model (for the same molecular 
composition) agrees with that of the self-consistent model. 

The fundamental significance of our parametric $P$-$T$ profile is the
ability to fit a disparate set of planetary atmosphere structures,
with very different atmospheric conditions (Figure~\ref{fig:fit_profs}). In
particular, the published hot Jupiter $P$-$T$ profiles were
obtained from several different modeling schemes. The different
planetary atmospheres all conform to a basic mathematical model
because of the common physics underlying the $P$-$T$ profiles. We anticipate 
that our parametric $P$-$T$ profile will open up a new avenue in
retrieving $P$-$T$ structure of exoplanetary atmospheres by enabling an
efficient exploration of the temperature structures and compositions
allowed by the data.

An important point concerns the adaptability of the model proposed
in this work. The introduction of the isothermal Layer 3 in 
(\ref{eqn:pt_profile}) is motivated by the fact that hot Jupiters,
which are the prime focus of this work, have convective-radiative
boundaries that are deep in the atmosphere, and the presence of 
an isothermal Layer 3 is physically plausible (as is 
evident from self-consistent calculations in literature). However, the 
model can be adapted to cases of cooler atmospheres where 
the convective-radiative boundary occurs at lower pressures 
than for hot Jupiters, and where there may not be an isothermal 
layer. The appropriate model $P$-$T$ profile in such cases would 
be one with only the two upper layers (Layer 1 and Layer 2). As 
can be seen from Figure~\ref{fig:fit_profs}, model fits to the solar system planets 
belong to the category of models with only Layers 1 and 2. 

\section{Model Atmosphere}

The goal of our model atmosphere is to be able to generate atmospheric spectra 
for a given planet over wide regions of parameter space and determine contours 
of fits with data. To meet this goal, our model uses efficient parametrization of 
the pressure-temperature structure and chemical composition of a 
planetary atmosphere. Our model atmosphere includes a 1D parametric $P$-$T$ 
profile and parametric molecular abundances, coupled with line-by-line radiative 
transfer, hydrostatic equilibrium, and the requirement of energy balance at the top 
of the atmosphere. Given a set of parameter values, the output is the emergent 
spectrum, and a goodness-of-fit to a given data set.

The major difference of our model from traditional atmosphere 
models is in the treatment of energy balance. Our model requires 
energy balance at the top of the atmosphere, instead of an 
iterative scheme to ensure layer-by-layer radiative (or radiative + 
convective) equilibrium as is done in conventional models. 
For a given set of model parameter values, we require that the net energy 
output at the top of the atmosphere (obtained by integrating the emergent 
spectrum) balances the net energy input due to the 
incident stellar flux (see \S \ref{sec-ebalance}). Models which do not 
satisfy this requirement are discarded. By running a large number of 
($\sim 10^7$) models in the parameter space, and discarding those 
that do not satisfy the above requirement, we are left with a population 
of models that satisfy energy balance.

Our temperature and abundance retrieval method also differs 
from other retrieval methods (Swain et al. 2008b, 2009a; Sing et al. 2008). Our work 
differs fundamentally by: using a parametric $P$-$T$ profile, exploring a wide range in 
parameter space ($\sim 10^4$ $P$-$T$ profiles independent of existing models, 
and $\sim 10^7$ models including abundances), maintaining  hydrostatic equilibrium 
and global energy balance, and placing constraints on the day-night energy redistribution. 
We, therefore, consider our method to be a new step towards atmospheric temperature 
and abundance retrieval for extrasolar planets.

\subsection{Radiative Transfer and Hydrostatic Equilibrium}

We calculate emergent spectra using a one-dimensional line-by-line 
radiative transfer code, assuming LTE.  Our model atmosphere consists 
of 100 layers in the pressure range between $10^{-5}$ bar and $100$ bar, 
and uniformly spaced in $\log(P)$. The key aspect of our model is the 
parametrization of the pressure-temperature profile and the chemical 
composition. In each layer, the parametric $P$-$T$ profile (described in 
\S~\ref{sec-parametricPT}) determines the temperature and pressure. And, 
the density is given by the ideal gas equation of state. The radial distance is 
determined by the requirement of hydrostatic equilibrium. And, the equation 
of radiative transfer, assuming LTE and no scattering, is then solved along 
with the adopted molecular abundances (see \S~\ref{sec-molecules} below) 
to determine the emergent flux. The geometry and the radiative transfer 
formulation are modified appropriately for calculating transmission spectra.

Our model directly computes the thermal emission spectrum and indirectly 
accounts for stellar heating and scattered radiation. In our model, we assume 
that the effects of heating by stellar radiation are included in the allowed 
``shapes'' of the parametric $P$-$T$ profile. As described in \S~\ref{sec-fit_pt}, 
our parametric $P$-$T$ profile fits the $P$-$T$ profiles obtained from 
self-consistent models that directly include absorbed stellar radiation, including 
profiles with thermal inversions. 

One reason the approach of indirectly including stellar irradiation is valid is that the 
visible and infrared opacities are largely decoupled (Marley et al. 2002). For
example, Na, K, TiO, VO can have strong absorption at visible
wavelengths in brown dwarfs and exoplanets, yet are mostly negligible
absorbers in the infrared. H$_2$O dominates infrared absorption, and
CH$_4$ and CO are also spectroscopically active in the IR. Another
reason the indirect account for stellar irradiation via a parametric
$P$-$T$ profile is justified is the uncertain chemistry (via unknown
optical absorbers that in some cases cause thermal inversions) and
physics that is hard to account for directly in a self-consistent 
1D model. We note that it is not known whether the 
common assumption of radiative equilibrium in self-consistent models, 
while one possible way of including stellar irradiation, holds for hot Jupiter 
atmospheres.  

We include molecules of hydrogen (H$_2$), water vapor (H$_2$O), carbon
monoxide (CO), carbon dioxide (CO$_2$), methane (CH$_4$), and ammonia
(NH$_3$). Our H$_2$O, CH$_4$, CO and NH$_3$ molecular line data are from 
Freedman et al. 2008, and references therein. Our CO$_2$ data are from 
Freedman (personal communication, 2009) and Rothman et al. (2005). And, we obtain 
the H$_2$-H$_2$ collision-induced opacities from Borysow et al. (1997), and 
Borysow (2002).

\subsection{Molecular Abundances}
\label{sec-molecules}
Molecular abundances in exoplanet atmospheres may depart from chemical
equilibrium (Liang et al. 2003; Cooper \& Showman, 2006). Our code, 
therefore, has parametric prescriptions to allow for non-equilibrium 
quantities of each molecule considered.  We begin by calculating a 
fiducial concentration of each molecule in a given layer, based on 
the assumption of chemical equilibrium.  For chemical equilibrium we 
use the analytic expressions in Burrows and
Sharp (1999), and assume solar abundances for the elements. The
molecules we include are H$_2$O, CO, CH$_4$, CO$_2$ and NH$_3$. The
molecule CO$_2$ is thought to originate from photochemistry 
and no simple expression for equilibrium composition is available; 
we therefore choose an arbitrary fiducial concentration. We denote the 
concentration of each molecule by its mixing ratio, i.e number fraction with 
respect to molecular hydrogen

The concentration of each molecule in a given layer is calculated by 
multiplying the corresponding fiducial concentration by a constant 
factor specific to each molecule. The constant factor corresponding B
to each molecule is a parameter of the model. Therefore, corresponding 
to the four prominent molecules, $\rm{H_2O}$, $\rm{CO}$, $\rm{CH_4}$ and
$\rm{CO_2}$, in our model, we have four parameters: $f_{\rm{H_2O}}$, 
$f_{\rm{CO}}$, $f_{\rm{CH_4}}$ and $f_{\rm{CO_2}}$. For instance, $f_{\rm{H_2O}}$ 
is the ratio of the concentration of H$_2$O to the fiducial concentration of H$_2$O
in each layer. Thus, although the concentration of H$_2$O is different in each 
layer, it maintains a constant ratio ($f_{\rm{H_2O}}$) with the fiducial concentration 
in that layer. By varying over a range in these four parameters, we are essentially 
varying over a wide range of molecular mixing ratios and elemental abundances. 

For NH$_3$, we restrict the concentration to the fiducial equilibrium concentration, 
i.e $f_{\rm{NH_3}} = 1$. This is because, NH$_3$ has limited and weak spectral 
features in the {\it Spitzer} and {\it HST} bands under consideration in this work.
Additionally, NH$_3$ is not expected to be abundant in enhanced quantities 
at such high temperatures as are seen in hot Jupiters. 

Clouds are not included in our model because HD 209458b and HD 189733b
likely do not have clouds that scatter or emit at thermal infrared wavelengths.
Several reasons have been proposed in literature justifying the use of cloud-free 
models for these planets: finding weak effects of clouds
on the P-T profile (Fortney et al. 2006 \& 2008); fast sedimentation rates in 
radiative atmospheres (Barman et al. 2005); clouds forming too deep in the 
planetary atmosphere to affect the emergent spectrum (Fortney et al. 2006); 
spectral features in transmission dominated by molecular absorption 
over cloud particles (Showman et al. 2009 and references therein); 
haze absorption at blue wavelengths on HD~189733b drops off rapidly 
with increasing wavelength and should be negligible at IR wavelengths (Pont et al. 2008). 
That said, clouds can be included in our model as another free parameter, via
a wavelength-dependent opacity, and we plan to make this extension in
the future. 

\subsection{Energy Balance}   
\label{sec-ebalance}

A planetary atmosphere model must satisfy the fundamental constraint
of energy conservation (here called energy balance). We consider a 
1D plane-parallel atmosphere with normal incidence. While using a 
plane-parallel model, it is imperative to weight the incident stellar flux 
appropriately in order to represent an average day side atmosphere. 

The average stellar flux incident on the planetary day side is given by 
$F_s = {\it f} F_\star$ where, $F_\star$ is the wavelength integrated 
stellar flux at the sub-stellar point of the planet. {\it f} is a geometric 
factor by which the stellar flux at the sub-stellar point must be weighted so 
as to represent an average 1D plane-parallel incidence. A value of ${\it f} = 
\frac{1}{2}$ represents uniform distribution of stellar flux on the planet 
day side; it comes from $F_\star \times (\pi R_p^2)/(2\pi R_p^2)$, i.e., the 
ratio of the planetary surface projected perpendicular to the incident 
stellar flux to the actual area of the planetary surface receiving the flux. 
However, it has been shown that just before secondary eclipse the dayside 
average flux should be biased towards a higher value of ${\it f} = \frac{2}{3}$ 
(Burrows et al. 2008). Therefore, in the present work, we adopt a value of 
${\it f} = \frac{2}{3}$. 

In the context of our 1D atmosphere, the constraint of energy balance 
means that the wavelength integrated emergent flux at the top of the planetary 
atmosphere matches the wavelength integrated incident stellar flux, after accounting
for the Bond albedo ($A_B$) and possible redistribution of energy onto the night 
side ($f_r$). Our prescription for the day-night redistribution is similar to
that of the $P_n$ prescription of Burrows et al. (2006). We define $f_r$ as 
the fraction of input stellar flux that is redistributed to the night side. 
The input stellar flux is given by $F_{in} = (1-A_B)F_s$, i.e., the incident stellar
flux ($F_s$), less the reflected flux ($A_BF_s$). Therefore, the energy 
advected to the night side is given by $f_r F_{in}$. And hence, the 
amount of flux absorbed on the dayside is given by $F_{in,{\rm day}} = (1-f_r) F_{in}$,
which equals $(1-A_B)(1-f_r)F_s$. It is assumed that the intrinsic flux from 
the planet interior is negligible compared to the incident stellar irradiation. 

Thus, the energy balance requirement on the day side spectrum is given as:
\begin{equation}
F_p = (1-A_B)(1-f_r)F_s,
\label{eqn:ebalance}
\end{equation}
where, $F_p$ is the wavelength integrated emergent flux calculated from
the model spectrum. We use a Kurucz model for the stellar spectrum (Castelli \& Kurucz, 2004). 
We adopt the stellar and planetary parameters from Torres et al (2008). 

We use energy balance, equation (\ref{eqn:ebalance}), in two ways. The first, as 
described above, is to ensure that our model atmosphere satisfies energy 
conservation. The constraint on the model comes from the fact that $A_B$ 
and $f_r$ are bounded in the range [0,1].  Thus, models for which $\eta =
(1 - A_B)(1 - f_r)$ is greater than unity are discarded on grounds of energy 
balance. Secondly, for models which satisfy energy balance, equation (\ref{eqn:ebalance}) 
gives us an estimate of $\eta$ for a given model. If an $A_B$ is 
assumed, this gives a constraint on $f_r$, and vice-versa.

\subsection{Parameter Space} 

The parameter space of our model consists of $N = N_{profile} +
N_{molec}$ free parameters, where $N_{profile}$ is the number of
parameters in the $P$-$T$ profile, and $N_{molec}$ is the number of
parameters pertaining to the chemical composition. 
As described in \S~2, our parametric $P$-$T$ profile 
has $N_{profile} = 6$, the six parameters being $T_0$, 
$\alpha_1$, $P_1$, $P_2$, $\alpha_2$, and $P_3$. And, corresponding to
the four prominent molecules, $\rm{H_2O}$, $\rm{CO}$, $\rm{CH_4}$ and
$\rm{CO_2}$, in our model, we have $N_{molec} = 4$ parameters:
$f_{\rm{H_2O}}$, $f_{\rm{CO}}$, $f_{\rm{CH_4}}$ and
$f_{\rm{CO_2}}$ (see \S\ref{sec-molecules}). Thus, our model 
has ten free parameters. 

Given a set of observations of a planet, for example a day-side
spectrum, our goal is to determine the regions of parameter space
constrained by the data, and to identify the degeneracies therein. We
investigate models with and without thermal inversions. To this
effect, we run a large number of models on a grid in the parameter
space of each scenario. Each parameter in either scenario has a finite
range of values to explore, and we have empirically determined the
appropriate step size in each parameter. A grid of reasonable
resolution in this ten-parameter space can approach a grid-size that
is computationally prohibitive. 

Even before running the grid of models, however, the parameter space 
of allowed models can be constrained to some extent based on the data 
and energy balance. Firstly, the model planet spectrum is bounded by 
two blackbody spectra, corresponding to the lowest and highest
temperatures in the atmosphere. Correspondingly, the maximum and
minimum temperatures, $T_{max}$ and $T_{min}$, of the $P$-$T$ profile are
partly constrained by the maximum and minimum blackbody temperatures
admissible by the data, and allowing a margin accounting for the fact that 
observations are available only at limited wavelengths. Secondly, an upper
limit can be placed on the temperature at the base of our model 
atmosphere ($T_3$ in the $P$-$T$ profile) based on energy balance. 
The maximum of $T_3$ is given 
by the effective temperature of the planet day-side assuming zero
albedo and zero redistribution of the energy to the night-side, and
allowing for a factor close to unity ($\sim 1 - 1.5$) to take into 
account the changes in the blackbody flux due to line features. 
Therefore, given the data, the space of $P$-$T$ profiles can be 
constrained to some extent even before calculating the spectra, 
using the above conditions.

The final grid is decided by running several coarse grids, and
obtaining an empirical understanding of the parameter space. At the
end, a nominal grid in ten parameters, with or without 
inversions in the $P$-$T$ profiles, consists of $\sim 10^7$ models. 
We thus computed a total of $\sim 10^7$ models for each planet
under consideration. We ran the models on $\sim 100$ 2-GHz 
processors on a Beowulf cluster, at the MIT Kavli Institute for 
Astrophysics and Space Research, Cambridge, MA.

\subsection{Quantitative Measure of Error} 
\label{sec-xisqr}
Broad-band observations of exoplanet atmospheres are currently limited 
to the six channels of broadband infrared photometry using 
\textit{Spitzer}. Additionally, a few narrow band spectrophotometry have also 
been reported using the {\it HST} NICMOS G206 grism (1.4 - 2.6 $\micron$), 
and the \textit{Spitzer} IRS spectrograph (5-14 $\micron$). The
number of broadband observations available in a single data set 
are typically smaller than the number of parameters in our model. 
Also, given that the parameters in our model are highly correlated 
and the degeneracies not well understood, the nature of existing 
data does not allow spectral fitting in the formal sense.

Consequently, our approach in this work is to compute a large number
of models on a pre-defined grid in the parameter space. For each
model, we calculate a ``goodness-of-fit" with the data using a
statistic defined by a weighted mean squared error, given by: 
\begin{equation}
\xi^2 = \frac{1}{N_{obs}} \sum_{i = 1}^{N_{obs}} \bigg(\frac{f_{i,model} - f_{i,obs}}
{\sigma_{i,obs}} \bigg)^2.
\label{eq:xi}
\end{equation}
Here, $f_i$ is the flux observable and $N_{obs}$ is the number of 
observed data points. $f_{i,obs}$ and $f_{i,model}$ are the observed $f_i$ and
the corresponding model $f_i$, respectively. For secondary eclipse
spectra, $f_i$ is the planet-star flux contrast, whereas, for
transmission spectra, $f_i$ is the transmission. $\sigma_{i,obs}$ is the 1
$\sigma$ measurement uncertainty in the observation. In order to
obtain the model fluxes in the same wavelength bins as the
observation, we integrate the model spectra with the transmission
functions of the instruments. Where spectrophotometry is available, 
we additionally convolve our model spectra with the instrument
point-spread function. 

The goodness-of-fit ($\xi^2$), as defined in (\ref{eq:xi}),  is similar 
in formulation to the conventional definition of the reduced $\chi^2$ 
statistic, if $N_{obs}$ is replaced by the number of degrees of 
freedom. In the current context, we refrain from using the $\chi^2$ statistic 
to asses our model fits because our number of model parameters ($N$) are 
typically more than the number of available broadband data points, leading 
to negative degrees of freedom. A value of $\chi^2$ with negative degrees 
of freedom lacks meaning in that we cannot relate it to a confidence level. 
Therefore, in using $\xi^2$ as our statistic of choice, we are essentially 
calculating $\chi^2$ per data point as opposed to $\chi^2$ per degree 
of freedom. We discuss this further in \S~\ref{sec-discuss-chisqr}. 

\section{Results: HD 189733b}

\subsection{Secondary Eclipse}
\label{sec-hd189}

HD~189733b has the best data set for any hot Jupiter known to date,
including both signal-to-noise and wavelength coverage.  Several
remarkable observations of the planet day side have been made by
determining the planet-star flux ratios just before secondary eclipse.
Grillmair et al. (2008) reported an infrared spectrum of the dayside in 
the 5 $\micron$ -- 14 $\micron$ range, obtained with the {\it Spitzer} 
IRS instrument. Charbonneau et al. (2008) presented broadband 
photometry in the six {\it Spitzer} channels (3.6, 4.5, 5.8, 8, 16, and 24 
$\micron$)\footnote{The 16 $\micron$ value of Charbonneau et al. (2008) 
was obtained by reanalyzing the observations of Deming et al. (2006), 
and has been found to be consistent with the latter.}. Separate photometric 
observations in the {\it Spitzer} 8 $\micron$ and 24 $\micron$ channels were 
also reported by Knutson et al. (2007) and (2009b), respectively. Additionally, 
an upper limit was placed on the 2.2 $\micron$ flux constrast by Barnes et al (2007).
More recently, Swain et al. (2009a) observed the planet dayside with 
{\it HST} NICMOS spectrophotometry in the 1.5 $\micron$ -- 2.5 $\micron$ range. 

\begin{figure*}[t]
\begin{center}
\includegraphics[width=\textwidth]{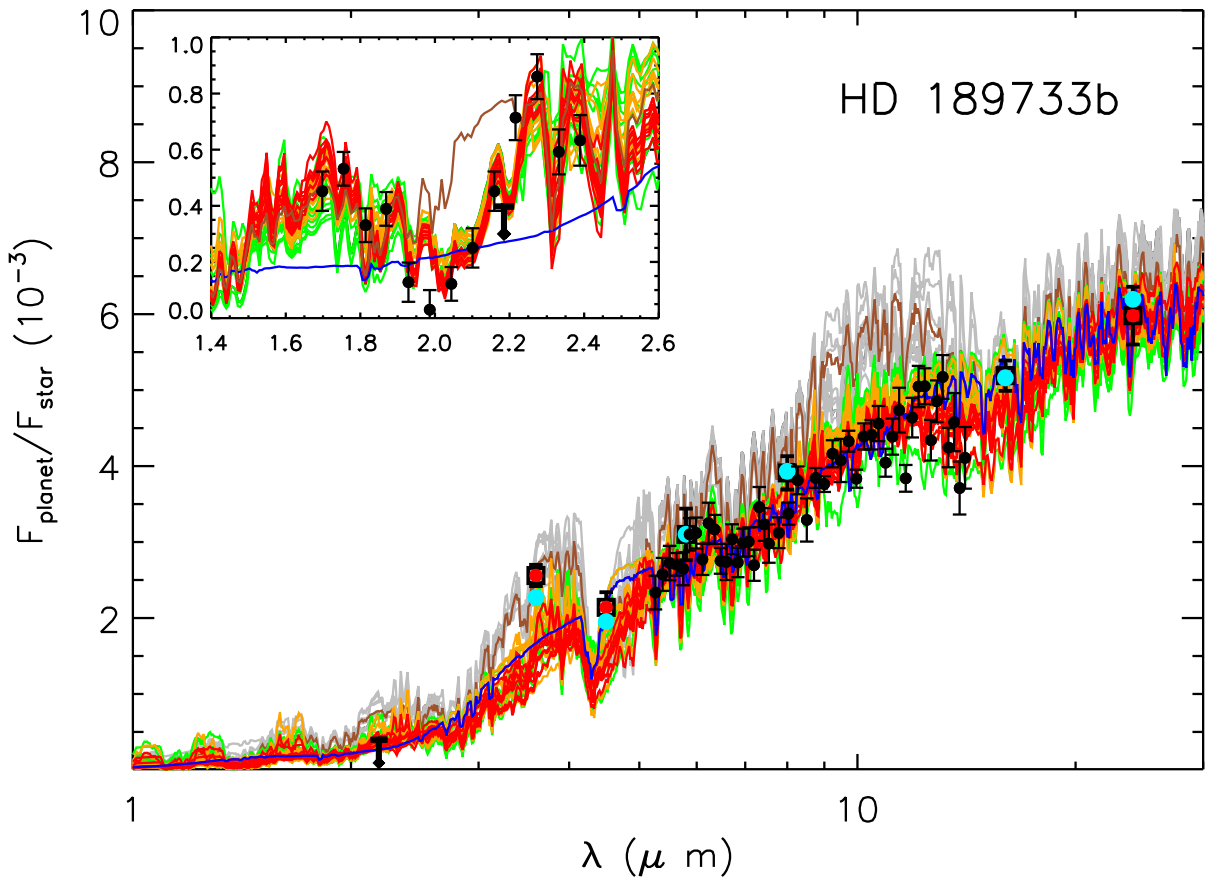}
\caption{Secondary eclipse spectra for HD 189733b. 
The black squares (filled with red) with error bars show the data from {\it Spitzer} 
photometry in six channels, at 3.6, 4.5, 5.8, 8, 16, and 24 $\micron$, 
reported by Charbonneau et al. 2008 (the 16 $\micron$ 
point is from reanalysis of Deming et al. 2006, consistent with the latter). The black circles with 
error bars in the main panel show the IRS data from Grillmair et al. 2008, and those in 
the inset show the {\it HST} NICMOS data from Swain et al.~2009a (see 
\S~\ref{sec-hd189} for details). The upper-limit at 2.2 $\micron$ shows the 
1-$\sigma$ constraint from Barnes et al. (2007). In the main panel,
the red, orange, and green spectra show model spectra that fit 
the IRS spectrum with $\xi^2$ in the range of 1.0 - 2.0, 2.0 - 3.0, and 
3.0 - 4.0, respectively. Ten spectra from each category are shown. 
For guidance, a single best fit model spectrum for this data is shown in blue. 
The gray spectra in the main panel show ten of the models that fit the {\it Spitzer} 
broadband photometry with $\xi^2 \leq 2.0$. For guidance, 
a single best fit model spectrum for this data is shown in brown (with the bandpass averaged points 
shown in light-blue). The inset shows ten of the model spectra that fit 
the NICMOS data; with colors following those in the main panel. The blue and 
brown models in the inset are one each best-fit to the IRS data and 
photometric data, respectively, showing that the best-fit models for the
IRS data and the broadband photometry do not match the NICMOS data.} 
\label{fig:spec_hd189}
\end{center}
\end{figure*}

\begin{figure*}[]
\centering
\includegraphics[width = 6in, height = 7in]{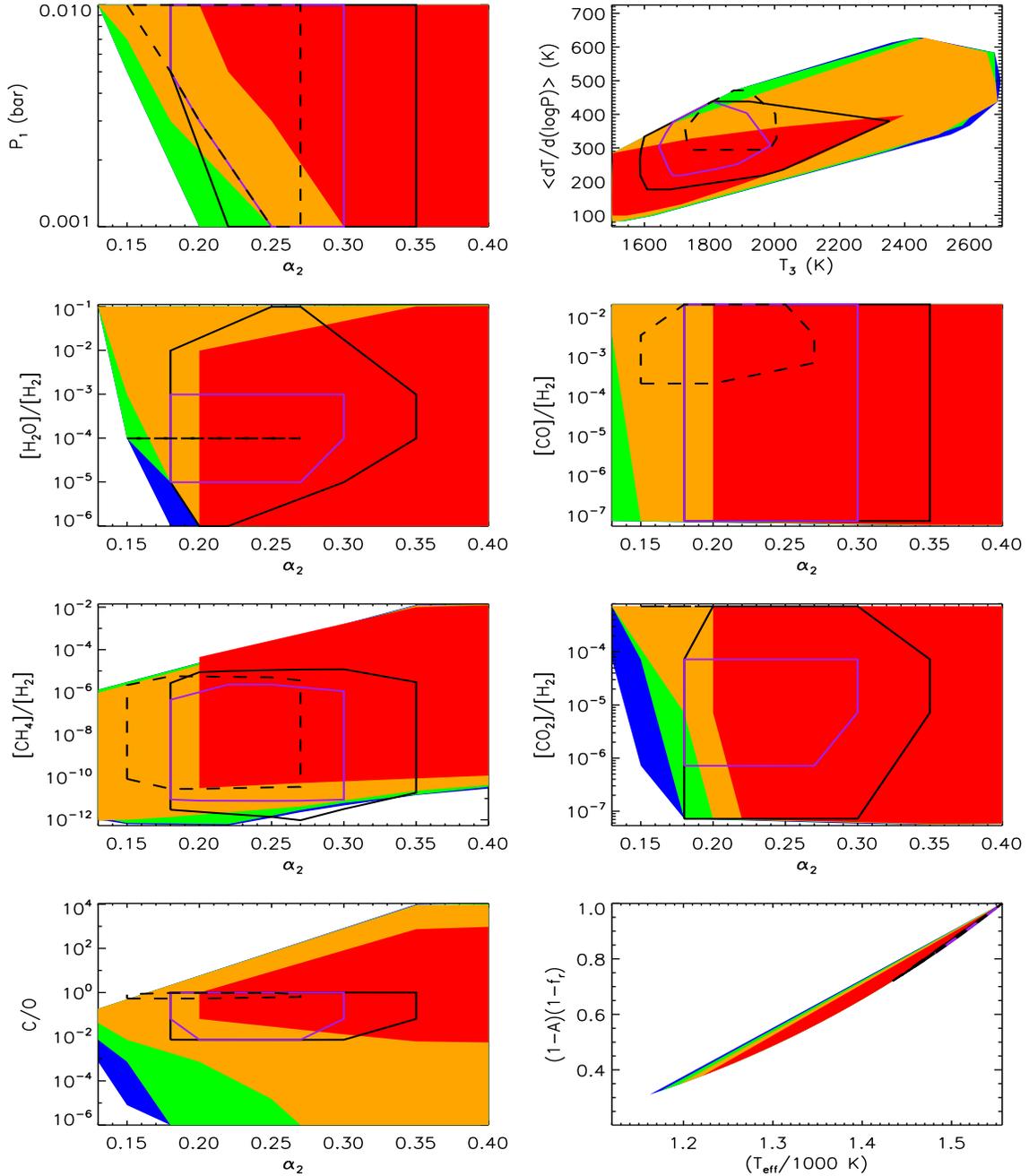}
\caption{Constraints on the dayside atmosphere of HD~189733b. 
The contours show minimum $\xi^2$ surfaces in the parameter space; 
at each point in a two-parameter space, 
the color shows the minimum possible $\xi^2$ out of a degenerate set of solutions
(see \S~\ref{sec:hd189_molec} for details). The filled color surfaces show the constraints
due to the {\it Spitzer} IRS spectrum (Grillmair et al. 2008). The red, 
orange, green and blue colors correspond to $\xi^2$ ranges 
of 1.0 - 2.0, 2.0 - 3.0, 3.0 - 4.0, and 4.0 - 5.0, respectively. The solid 
lines show the constraints from  {\it Spitzer} broadband photometry 
(Charbonneau et al. 2008); the black (purple) solid lines correspond to
the $\xi^2 < 2$ ($\xi^2 < 1$) surface. The dashed lines show the $\xi^2 < 2$ 
constraints from  {\it HST} NICMOS Spectro-photometry (Swain et al. 2009a). 
The results follow from $\sim 9 \times 10^6$ models. The mixing ratios of 
species are shown as ratios by number. $<{\rm dT/d(log P)}>$ is the 
average temperature gradient in Layer 2 of the P-T profile, which contributes 
most to the emergent spectrum.}

\label{fig:c_hd189}
\end{figure*}

\begin{figure}
\begin{center}
\includegraphics[width=0.5\textwidth]{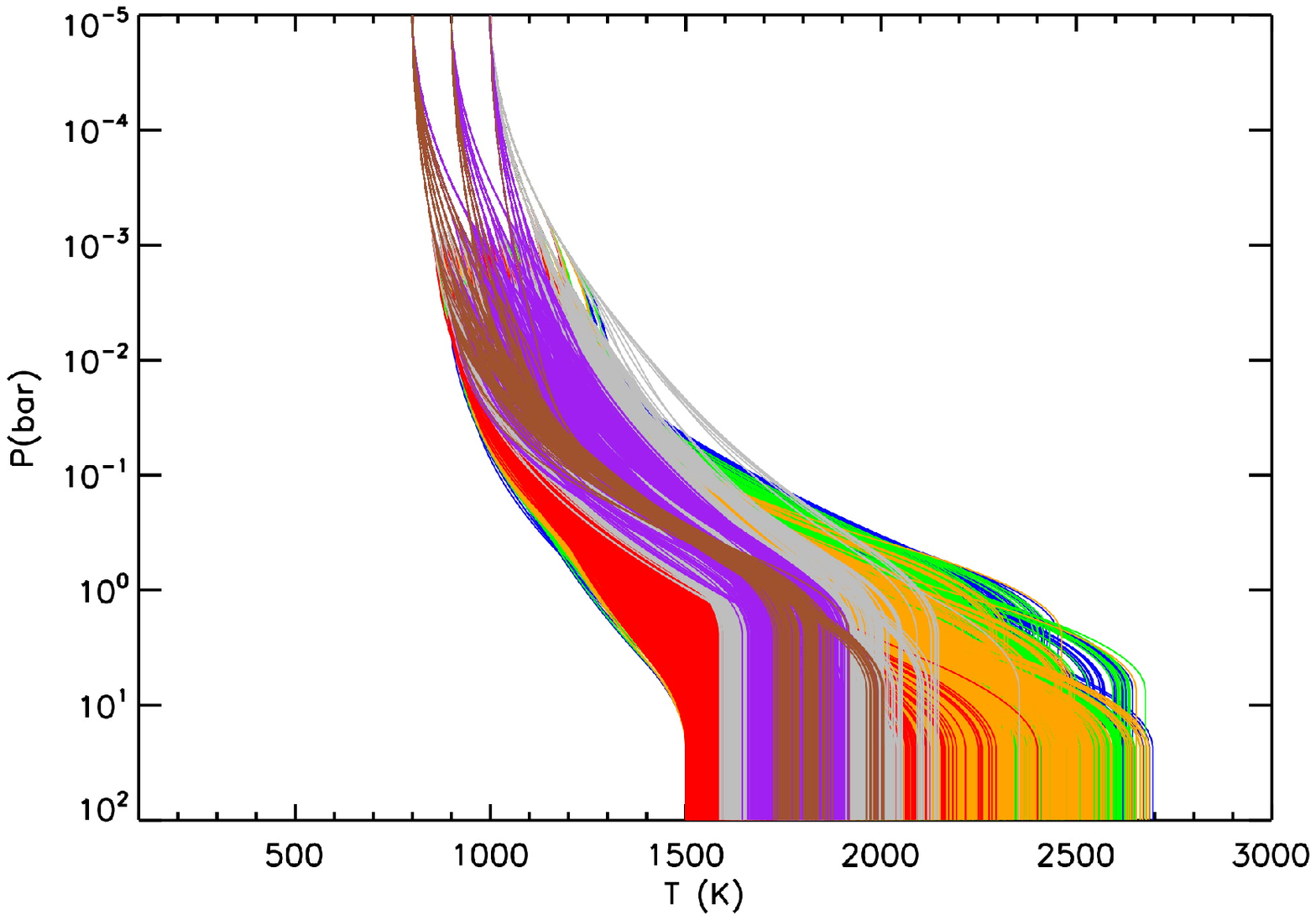}
\caption{Pressure-Temperature profiles explored by models for the secondary 
eclipse spectrum of HD 189733b. Since each $P-T$ profile can have multiple 
values of $\xi^2$, corresponding to the several possible molecular compositions, 
we color-code each $P-T$ profile with only the minimum possible value of $\xi^2$
(as also described in Figure~\ref{fig:c_hd189}). The red, orange, green and blue colors correspond to models that fit the IRS spectrum (Grillmair et al. 2008) with minimum $\xi^2$ in the ranges of 1.0 - 2.0, 2.0 - 3.0, 3.0 - 4.0, and 4.0 - 5.0, respectively. The purple and gray profiles correspond to models that fit the {\it Spitzer} broadband photometry  (Charbonneau et al. 2008) with minimum $\xi^2$ in the ranges 
0.0 - 1.0 and 1.0 - 2.0 respectively, and the brown profiles show models that 
fit the NICMOS data (Swain et al. 2009a) with the minimum $\xi^2$ in the 1.0 - 2.0 range.}
\label{fig:pt_hd189}
\end{center}
\end{figure}

Together, these data sets place important constraints on the dayside atmosphere
of the planet. The high S/N IRS spectrum reported by Grillmair et al (2008), 
obtained with 120 hours of integration time, is the best exoplanet spectrum 
known. This spectrum forms the primary data set for our present 
study of HD~189733b. In addition, we also use the photometric observations
of Charbonneau et al. (2008), and the spectrophotometric observations 
of Swain et al.~(2009a) to place complementary constraints \footnote{We do not 
use those binned data points of Swain et al.~(2009a) that are either at the edges 
of the chip or are non-detections at the 3-$\sigma$ level. These include points 
with $\lambda < 1.65~\micron$ and  $\lambda > 2.4~\micron$}. We use the two 
independent observations of 
Knutson et al. (2007 \& 2009b), at 8 $\micron$ and 24 $\micron$, only as a guide 
for the models, and exclude them from our fits. This is because we place constraints from each 
individual data set separately, for reasons explained below, and we 
cannot place any constraints from a single data point from each of 
Knutson et al. (2007 \& 2009b). However, it must be noted that the 8 $\micron$ 
and 24 $\micron$ values of Knutson et al. (2007 \& 2009b) agree with the 
IRS spectrum and the Charbonneau et al. (2008) observations, respectively, 
which are being used in the current work. We present the data in Figure~
\ref{fig:spec_hd189}, along with model fits. 

We analyze each data set separately. The reason we use a separate 
treatment is that some individual observations at similar wavelengths differ from each 
other. Specifically, the 8 $\micron$ {\it Spitzer} IRAC data point from Charbonneau 
et al (2008) is $\sim 2 \sigma$ above the IRS spectrum at the same wavelength 
(Grillmair et al. 2008). Also, 8 $\micron$ IRAC observations taken at separate 
times differ by over 2 $\sigma$ (Charbonneau et al. 2008). From a model interpretation 
viewpoint, a 2$\sigma$ spread means that almost no constraints on molecular abundances 
can be drawn (based on the reported uncertainities in the data). In addition, the very strong 
evidence of CO$_2$ that is visibly obvious around 2~$\micron$ in the HST/NICMOS 
data set is not so strong in the 4.5~$\micron$ and 16~$\micron$ {\it Spitzer} channels. 
In this work, we are unable to find models that fit all the three data sets of HD~189733b 
simultaneously within the $\xi^2 = 2$ level. The data and model results below may 
well indicate atmospheric variability in this planet. See \S~\ref{sec-variability} for further 
discussion about possible atmospheric variability.

In what follows, we describe the constraints on the atmospheric properties 
of HD~189733b placed by each data set. Ideally, we would like to report 
constraints at the $\xi^2 = 1$ surface, i.e. only those models which, on an 
average, fit the observations to within the 1$\sigma$ error bars. However, we 
adopt a more conservative approach and lay emphasis on the results at 
the $\xi^2 = 2$ surface as well, allowing for models fitting the observations to 
within $\sim 1.4 \sigma$ on an average. The latter choice is somewhat 
arbitrary, and the informed reader could choose to consider other $\xi^2$ 
surfaces; we show contours of $\xi^2$ in the different atmospheric properties.

\subsubsection{Molecular Abundances}
\label{sec:hd189_molec}
The constraints on the abundances due to the different data sets are 
shown in Figure~\ref{fig:c_hd189}. The figure shows contours of $\xi^2$ in 
the parameter space of our model atmosphere. The contours show minimum $\xi^2$ surfaces. 
At each point in a given two-parameter space, several values of $\xi^2$ are possible 
because of degeneracies with other parameters in the parameter space. 
In other words, the same point in a pair of parameters can have multiple 
values of $\xi^2$ corresponding to evaluations at various values in the remaining parameters. 
For the contours in Figure~\ref{fig:c_hd189}, we consider the minimum possible 
value of $\xi^2$ at each point in a two-parameter space. 
Consequently, each colored surface shows the region of parameter 
space which allows a  minimum possible $\xi^2$ corresponding to that color.
The mixing ratios of species are shown as ratios by number.

{\it Spitzer IRS spectrum:} We begin with the IRS spectrum (Grillmair
et al.  2008) because this data set has the broadest wavelength
coverage and the best S/N. Also, the IRS data set has more data points
that the number of parameters in our model. Our most significant new 
finding is the existence of models fitting the IRS data set that are totally consistent
with efficient redistribution of energy from the planet dayside to the
night side (see Figure~\ref{fig:c_hd189} for the range of allowed
values). This addresses a significant problem in previously published
models for dayside spectra (Grillmair et al. 2008) not being able to
explain the efficient redistribution found in the phase curves of
Knutson et al. (2007 \& 2009b) (See Charbonneau et al. 2008, and Grillmair et
al. 2008; but also see Barman 2008). With the IRS data, we also
confirm the absence of a thermal inversion in HD 189733b. 

Despite the high S/N, the IRS spectrum allows only weak constraints 
on the molecular abundances. The reason being that the IRS spectral 
range has molecular absorption features of only CH$_4$ and H$_2$O;
even for these molecules, their features at other wavelengths remain 
completely unconstrained. The significant features are of CH$_4$ 
(7.6 $\micron$), and H$_2$O (6.3 $\micron$). For example, the 
$\xi^2 = 2$ surface constrains the mixing ratio of water vapor 
to $10^{-6} - 0.1$, and that of CH$_4$ to be less than $10^{-2}$ 
(but c.f \S~\ref{sec-model_ext}).

{\it Spitzer broadband photometry:} Our model fits to the {\it Spitzer} 
broadband data (Charbonneau et al. 2008) require less efficient 
circulation of incident stellar energy compared to the best model fits 
to the {\it Spitzer} IRS spectrum. Specifically, the value for redistribution 
in our framework is $f_r \lesssim 0.25$, and this agrees with previous results  
(Barman 2008, Charbonneau et al. 2008) which find that model
fits to the broadband data require inefficient day-night redistribution 
of energy. The discrepancy in redistribution arises from different planet-star 
contrast ratios. 
The 8 $\micron$ IRAC point of Charbonneau et al  (2008) is 
$\gtrsim  2 \sigma$ different from the IRS spectrum, and 
from the 8 $\micron$ IRAC measurement of Knutson et al (2007). Similar to 
the IRS spectrum, the {\it Spitzer} broad band photometry shows no sign 
of a thermal inversion.

The {\it Spitzer} broadband photometry enables significant constraints on 
molecular abundances. Again, we take the $\xi^2 = 2$ surface, 
corresponding to an average difference of 1.4 $\sigma$ between the model 
and data. We find the lower limit of the H$_2$O mixing ratio to be $10^{-6}$,
and the upper limit of the CH$_4$ mixing ratio to be $10^{-5}$ 
(Figure~\ref{fig:c_hd189}). At the $\xi^2 = 2$ surface there are no constraints on 
the presence or abundances of CO and CO$_2$. 
At the $\xi^2 = 1$ surface, the best fit 
models constrain the molecular mixing ratios as follows:  
$10^{-5} \leq$ H$_2$O $\leq 10^{-3}$; CH$_4$ $\leq 2 \times 10^{-6}$; 
$7 \times 10^{-7} \leq$ CO$_2$ $\leq 7 \times 10^{-5}$; CO is 
unconstrained because it is degenerate with CO$_2$. The 
identification of CO$_2$, at the $\xi^2 = 1$ surface, is intriguing, 
and this is the first time CO$_2$ is identified with {\it Spitzer} photometry. 
Our model also finds the C/O ratio to be $7 \times 10^{-3} \leq {\rm C/O} \leq 1$ for both the 
$\xi^2 = 1$ and $\xi^2 = 2$ surfaces. 

{\it HST/NICMOS~Spectro-photometry:} The {\it HST}/NICMOS data (Swain et al. 2009a) are 
useful because the wavelength range includes significant features 
of H$_2$O, CH$_4$, CO, and CO$_2$ (Figure~\ref{fig:spec_hd189}). We can 
therefore constrain the abundances even at the $\xi^2 = 2$ surface 
(Figure~\ref{fig:c_hd189}). We find mixing ratios of: H$_2$O  $\sim 10^{-4}$; 
CH$_4$ $ \leq 6 \times10^{-6}$; $2 \times 10^{-4} \leq$ CO $\leq 2 \times 10^{-2}$; 
and CO$_2$ $\sim 7 \times10^{-4}$. It is remarkable that 
at the $\xi^2 = 2$ surface we can constrain the abundances so stringently. 
The exception is CH$_4$, where the  upper limit indicates that methane 
is not present in large quantities on the dayside of HD~189733b. The 
C/O ratio is between 0.5 and 1. 
Our inferred abundances are consistent with those reported in Swain et al. (2009a), 
with the major exception of CO$_2$. While we find 
CO$_2$ mixing ratios of $7 \times 10^{-4}$, Swain et al. (2009a) 
find  $10^{-7} \leq$ CO$_2$ $\leq 10^{-6}$. At present we have no 
explanation for this discrepancy. 

A mixing ratio of $7 \times 10^{-4}$ is seemingly high from a theoretical 
standpoint; photochemistry which is thought to be the primary source of CO$_2$ 
in hot Jupiters allows mixing ratios of CO$_2$ up to $\sim 10^{-6}$, assuming 
solar metallicities (Liang et al. 2003, Zahnle et al. 2009).
One possible explanation of a large inferred CO$_2$ mixing might be a  
high mixing ratio concentrated in only a few layers of the atmosphere. 
The data can be fit equally well with the same mixing ratio 
($7 \times 10^{-4}$) of CO$_2$ present only between $10^{-3} - 0.3$ 
bar causing an average mixing ratio of $10^{-6}$ over the entire 
atmosphere. 

\begin{figure*}
\begin{center}
\includegraphics[width=\textwidth]{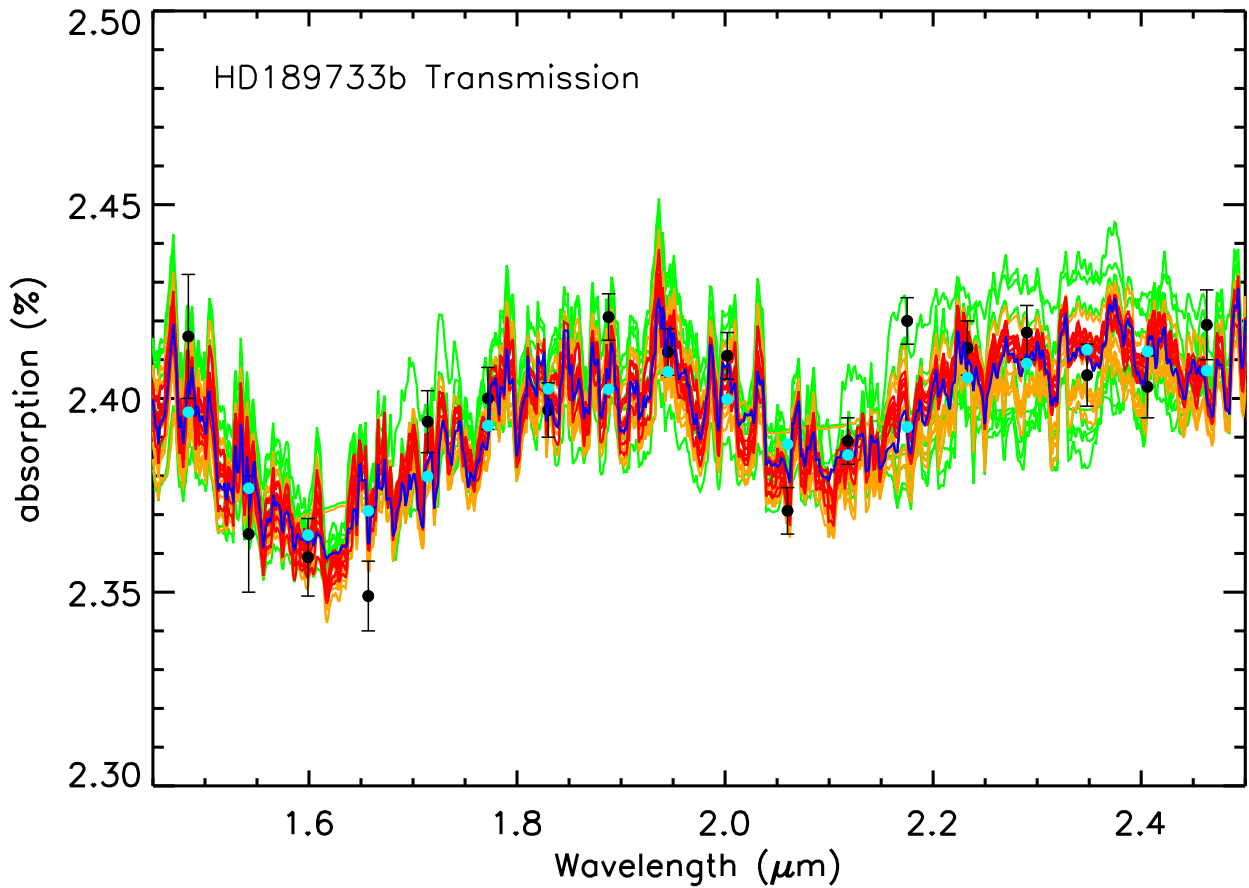}
\caption{Transmission spectra for HD~189733b. The black filled circles with error bars  show the data from Swain et al. 2008b (see \S~\ref{sec-hd189_trans} for details). 
The red, orange, and green spectra show ten models each in the $\xi^2$ ranges of 1.0 - 2.0, 2.0 - 3.0, and 3.0 - 4.0, respectively. The blue spectrum shows one best-fit model, with the light-blue filled circles showing the corresponding model points binned to the grism bandpass, after convolving with the instrument point spread function as described in Swain et al. (2008b).}
\label{fig:spec_hd189_trans}
\end{center}
\end{figure*}

\begin{figure*}[t]
\centering
\includegraphics{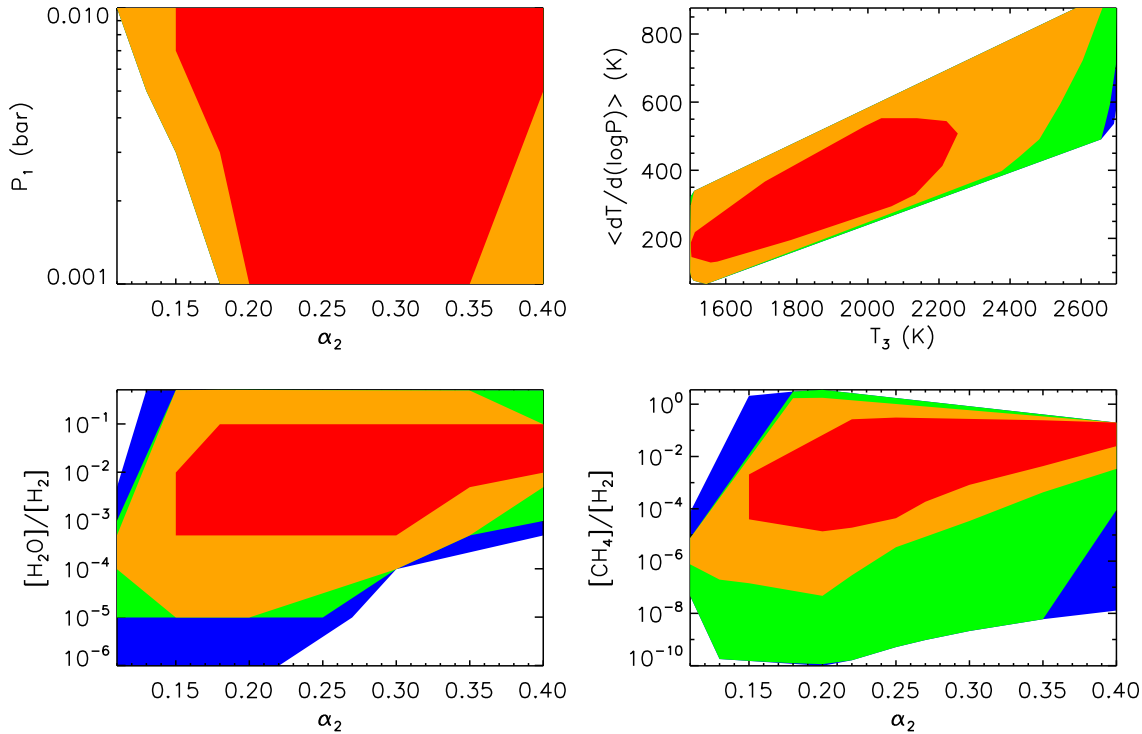}
\caption{Constraints on atmospheric properties at the limb of HD~189733b. 
The contours show minimum $\xi^2$ surfaces (see caption of Figure~\ref{fig:c_hd189}  for details), as constrained by the transmission spectrum of Swain et al. (2008b) 
described in \S~\ref{sec-hd189_trans}. The red, orange, green and blue surfaces  correspond to $\xi^2$ in the ranges 1.0 - 2.0, 2.0 - 3.0, 3.0 - 4.0, and 4.0 - 5.0, respectively.} 
\label{fig:c_hd189_trans}
\end{figure*}

\begin{figure}
\begin{center}
\includegraphics[width=0.5\textwidth]{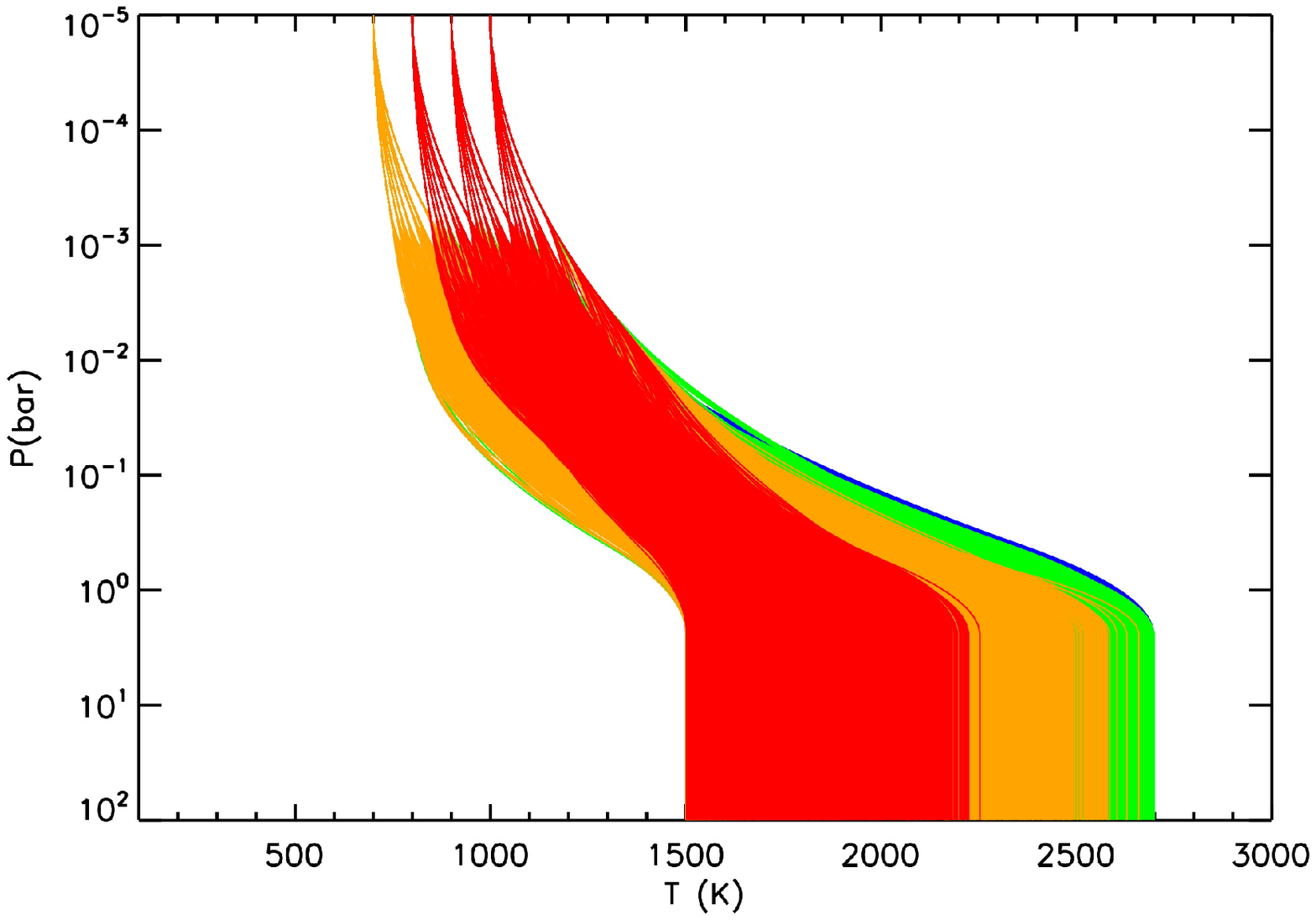}
\caption{Pressure-Temperature profiles explored by models for the
transmission spectrum of HD 189733b. The profiles are color-coded 
by minimum $\xi^2$ (see Figure~\ref{fig:pt_hd189}), and the colors 
are described in Figure~\ref{fig:c_hd189_trans}.}
\label{fig:pt_hd189_trans}
\end{center}
\end{figure}

\subsubsection{Pressure-Temperature Profiles}
Our results include constraints on the 1D averaged $P$-$T$ profile.
We explore the parameter space with a grid of $\sim 10^4$ $P$-$T$ profiles,
shown in Figure~\ref{fig:pt_hd189}. For $P$-$T$ profiles with no thermal inversions, the
parameters that influence the spectrum most are the two pressures
($P_1$ and $P_3$), which contribute the temperature differential in
layer 2 of the atmosphere, and the corresponding slope
($\alpha_2$). Additionally, $P_3$ governs the base flux, corresponding
to the blackbody flux of temperature $T_3$, on which the absorption
features are imprinted.

The constraints on the parameters of the $P$-$T$ profiles (\S~\ref{sec-parametricPT}) 
from the three data sets are shown in Figure~\ref{fig:c_hd189}. 
The IRS spectrum (Grillmair et al. 2008) constrains
the $P$-$T$ profiles towards larger values of $\alpha_2$, leading to
less steep temperature gradients.  For the same base 
temperature ($T_3$) of the planet dayside, the maximum temperature gradients allowed by the
broadband photometry and the {\it HST} spectrum are always
higher than that allowed by the IRS data set. The $\xi^2 = 2$ surface
for the IRS data set constrains $\alpha_2$ to be greater than 0.2,
while the constraints due to the broadband photometry and the {\it HST} data 
are 0.18 - 0.35 and 0.15 - 0.27, respectively. The same surface constrains the mean temperature
gradient ($<dT/d \log P>$) in layer 2 to 100 - 400 K, 180 - 440 K and
290 - 470 K for the IRS, the broadband photometry and the {\it HST} data sets, 
respectively. The $\xi^2 = 2$ surface also shows that the IRS data allows for 
lower base temperatures as compared to the other data sets. The skin temperature
($T_0$) at the top of the atmosphere (at $P \sim 10^{-5}$ bar) is relatively 
unconstrained because of its degeneracy with the temperature gradient, and 
because of the low contribution of spectral features at that pressure. 

\subsubsection{Albedo and Energy Redistribution}
\label{sec-hd189-fr}
Our results also constrain the albedo and day-night energy
redistribution.  The relevant parameter is $\eta = (1-A_B)(1-f_r)$, where 
$A_B$ is the Bond albedo, and $f_r$ is the fraction of input stellar 
flux redistributed to the night side (see \S~\ref{sec-ebalance}).  Since we can only
constrain the product $\eta$ and not the individual components, $A_B$
and $f_r$, one can estimate the constraint on $A_B$ given $f_r$ or
vice-versa. This is similar to other forward models in the literature, 
where an $f_r$ is assumed, and the albedo is determined 
by the model. Figure~\ref{fig:c_hd189} shows the $\xi^2$ surface in the space of
$T_{\rm eff}$ vs. $\eta$. The IRS data set allows for efficient day-night
circulation on this planet, with $\eta$ in the range 0.38 - 0.99 at the $\xi^2 = 2$ level. 
While $\eta = 1$ means zero albedo and no redistribution, a value 
of $\eta = 0.38$ translates into an upper-limit of $f_r \leq 0.62$ (for zero albedo).
Even if one were to assume $A_B = 0.3$ (a relatively high value for 
hot Jupiters; see Rowe et al. 2008), $f_r$ can be as high as 0.46, 
indicating very efficient redistribution. The broadband photometry and the {\it HST} data sets, on the 
other hand, require much lower efficiency in energy redistribution, 
constraining the lower bound on $\eta$, at the $\xi^2 = 2$ level, to values of 0.74 
and 0.72, respectively. These values translate into upper limits  on $f_r$ of 0.26 and 
0.28, respectively, suggesting low redistribution. The range of effective temperatures 
allowed by the different data sets follow accordingly. While the IRS data set allows for
$T_{\rm eff}$ between 1220 - 1550 K, the broadband photometry and {\it HST} 
data sets constrain $T_{\rm eff}$ between 1440 - 1560 K. 

It is natural to ask: what can we say with the combined data sets from {\it Spitzer} 
IRS, {\it Spitzer} IRAC, and {\it HST} NICMOS? The answer depends on which 
error surface one is willing to adopt. Each of the data sets places a different 
constraint on the planet atmosphere (at the $\xi^2 \leq 2$ surface). Thus, 
if we take the data sets at face value, and adopt the model $\xi^2 \leq 2$ surfaces,  
we find that the atmosphere must be variable, both in its energy redistribution state
and in the chemical abundances.  Specifically, the high CO$_2$ value could be 
pointing to local changes due to transient photochemical effects and atmospheric flows.
On the other hand, inference of stellar variability, or refined observational uncertainties 
may potentially change this situation.

\subsection{Transmission Spectra}
\label{sec-hd189_trans}

The transmission spectrum of HD189733b was presented by Swain et
al. (2008b), who reported the detection of CH$_4$ and H$_2$O in the 1.4
$\micron$ - 2.5 $\micron$ range covered by the {\it HST}/NICMOS
spectrophotometry. Transmission data are complementary to secondary
eclipse data, because a transmission spectrum probes the part of the 
planet atmosphere in the vicinity of the limb, whereas secondary
eclipse data probes the planetary day side. In addition, a transmission 
spectrum reveals line features of the species in absorption, imprinted 
on the stellar flux, with the self-emission of the planet contributing  
negligibly to the observed flux. 

We computed $\sim 7 \times 10^5$ models to place constraints on
allowed limb $P$-$T$ profiles and quantify the allowed abundances by the 
$\xi^2$ surfaces.  In order to compute the $\xi^2$, our models
were convolved with the instrument point-spread function, integrated
over the grism transmission function, and finally binned to the same
bins as the data (as described in Swain et al. 2008b). We find that 2
of 18 binned data points (at 1.888 $\micron$ and 2.175 $\micron$ ) as
reported by Swain et al. (2008b) have unusually high values, and 
amount to an unusually large contribution to the net $\xi^2$. 
In the framework of our models, we do not find any spectral features 
that might be able to explain those values. Consequently, we leave
these two points out of our present analysis. Figure~\ref{fig:spec_hd189_trans} 
shows the model spectra and data. 

\subsubsection{Molecular Abundances} 
Our results confirm the presence of H$_2$O and CH$_4$ in the limb, 
as also reported by Swain et al (2008b). At the $\xi^2=2$ surface, as shown 
in Figure~\ref{fig:c_hd189_trans}, the H$_2$O mixing ratio is constrained 
to $5 \times 10^{-4} - 0.1$, and the 
CH$_4$ mixing ratio lies between $10^{-5} - 0.3$. These ranges of 
mixing ratios for H$_2$O and CH$_4$ are consistent with the mixing 
ratios found for the best-fit model reported by Swain et al (2008b). Although, 
these constraints are roughly consistent with the constraints on the 
same molecules on the dayside inferred in \S~\ref{sec-hd189}, the 
high mixing ratios of CH$_4$ allowed is appaling at first glance. 
This hints at a region of high CH$_4$ concentration in the limb, 
possibly on the night side.  The effect of other possible molecules, 
CO, NH$_3$ and CO$_2$ on the spectrum is minimal. Although 
each of these molecules have some spectral features in 
the observed wavelength range, the upper-limits on their 
compositions placed by the data are not statistically significant.

\subsubsection{Pressure-Temperature Profiles}

Figure~\ref{fig:pt_hd189_trans} shows the $P$-$T$ profiles explored
for the HD~189733b transmission data set.  The best-fit models with $\xi^2 \leq 2$ are shown 
in red. Figure~\ref{fig:c_hd189_trans} shows the $\xi^2$ surface in 
the parameters space of the $P$-$T$ profiles. The best-fit models
are consistent with an atmosphere with no thermal inversions. The 
$\alpha_2$ parameter is constrained by the $\xi^2 = 2$ surface to be 
greater than $\sim 0.15$, and the temperature $T_3$ at the limb is 
constrained to lie between 1500 K - 2250 K. The mean temperature 
gradient ($<dT/d \log P>$) in layer 2 constrained by the $\xi^2 = 2$ surface 
lies between 130 K - 550 K.

\begin{figure*}
\begin{center}
\includegraphics[width=\textwidth]{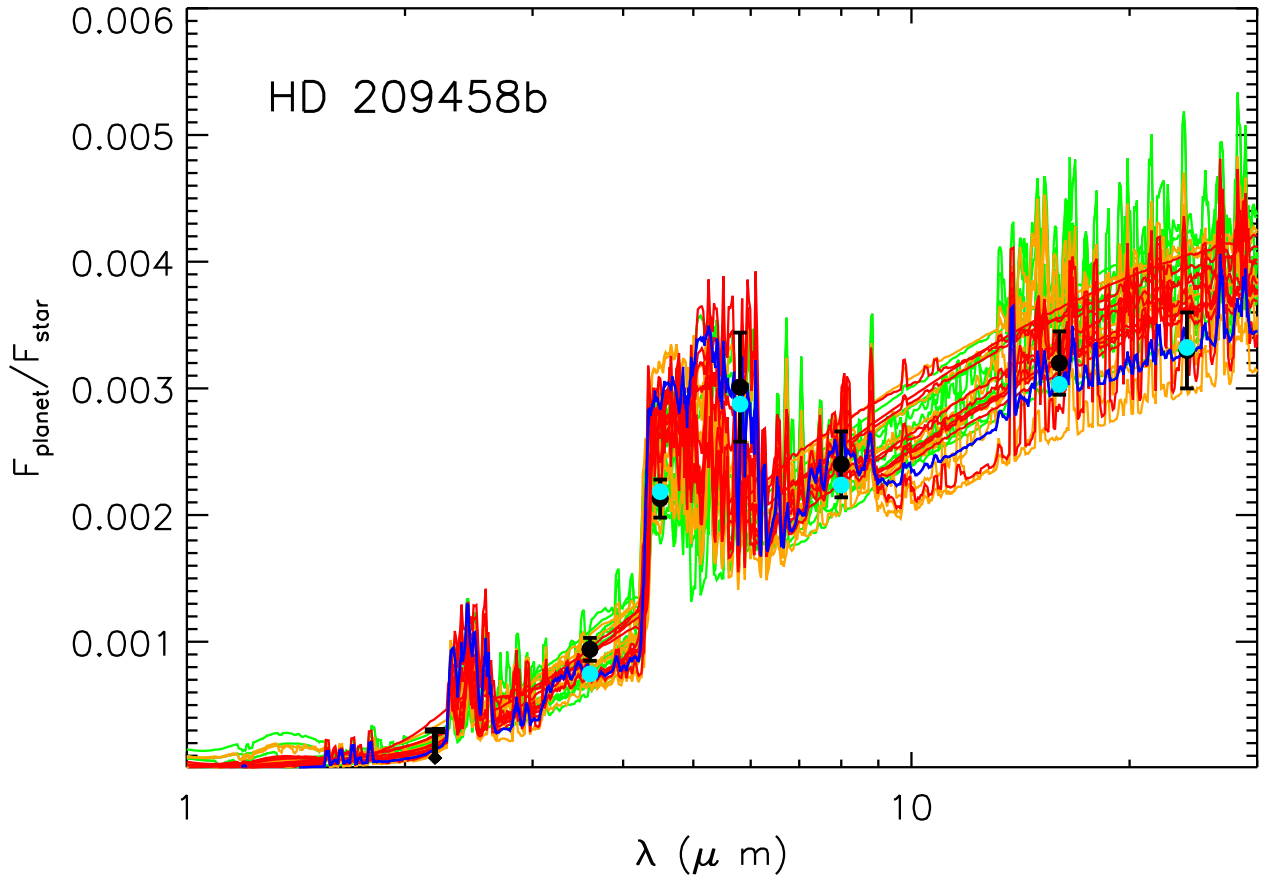}
\caption{Secondary Eclipse spectra for HD 209458b. 
The black filled circles with error bars show the data, obtained by 
{\it Spitzer} photometry. The 3.6 $\micron$, 4.5 $\micron$, 5.8 $\micron$, 
and 8 $\micron$ data are from Knutson et al. (2008). The 16 $\micron$
and 24 $\micron$ data are from Deming (personal communication, 2009). 
The upper-limit at 2.2 $\micron$ shows the 1-$\sigma$ constraint from Richardson 
et al. (2003). The red, orange, and green spectra correspond to models 
with $\xi^2$ in the range of 0.0 - 2.0, 2.0 - 3.0, and 3.0 - 4.0, respectively. 
For guidance, a single best fit model spectrum is shown 
in blue (with the bandpass averaged points shown in light-blue).}
\label{fig:spec_hd209}
\end{center}
\end{figure*}

\begin{figure*}[t]
\centering
\includegraphics[width = 6in, height = 7.5in]{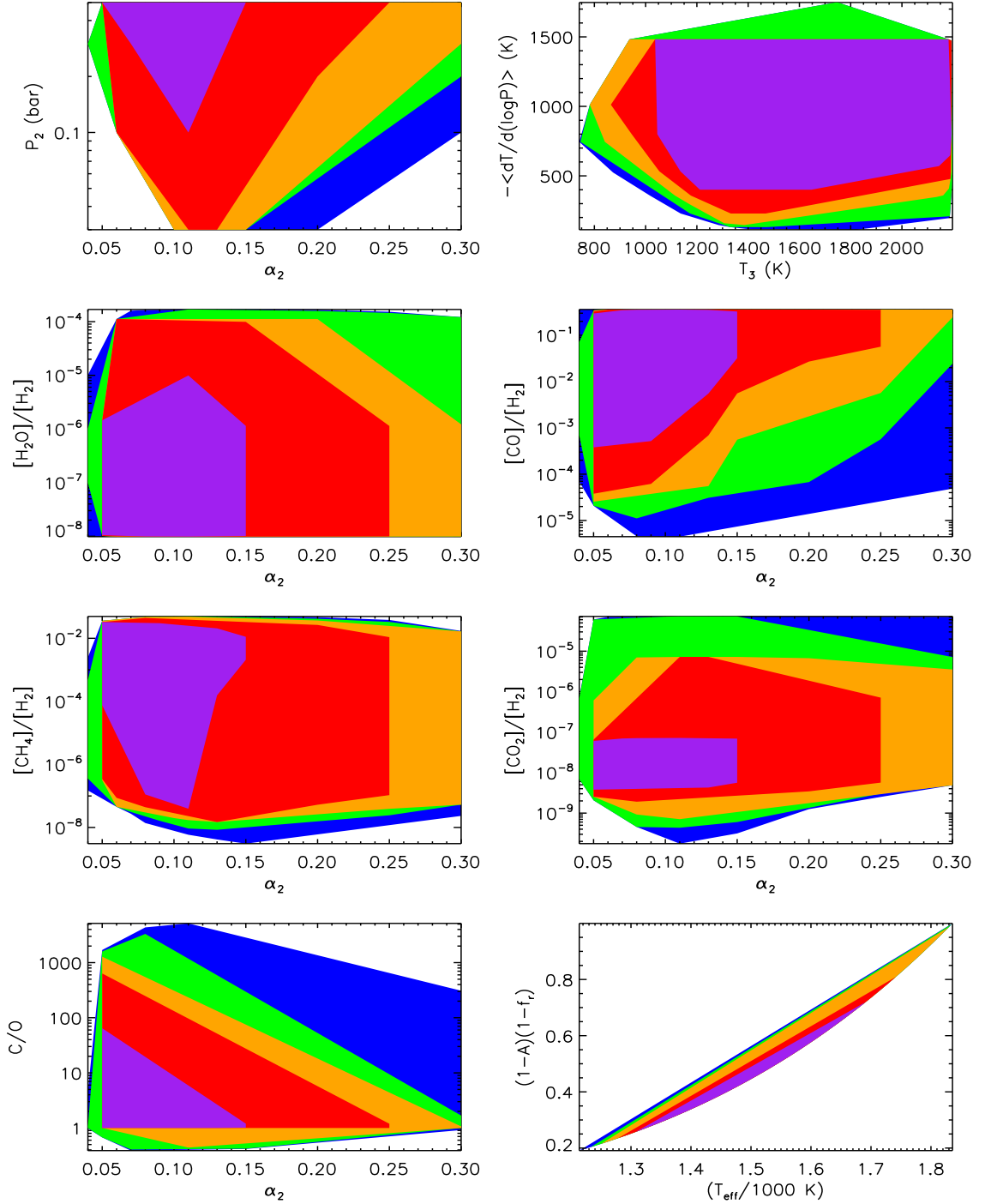}
\caption{Constraints on the dayside atmosphere of HD 209458b.
The contours show minimum $\xi^2$ surfaces (see Figure~\ref{fig:c_hd189}) 
in the parameter space of models for secondary eclipse spectra of HD 209458b. 
The red, orange, green and blue surfaces  correspond to $\xi^2$ in the ranges 
1.0 - 2.0, 2.0 - 3.0, 3.0 - 4.0, and 4.0 - 5.0, respectively. Additionally, 
the purple colored surfaces show regions that allow $\xi^2 < 1$. The 
results follow from $\sim 6 \times 10^6$ models with thermal inversions. 
Models without thermal inversions are ruled out by the data.  
$<{\rm dT/d(log P)}>$ is the average temperature gradient in the 
atmosphere between pressures P$_1$ and P$_2$, in $P$-$T$ profiles 
with thermal inversions, which contributes most to the observed emission 
features.}
\label{fig:c_hd209}
\end{figure*}

\begin{figure}
\begin{center}
\includegraphics[width=0.5\textwidth]{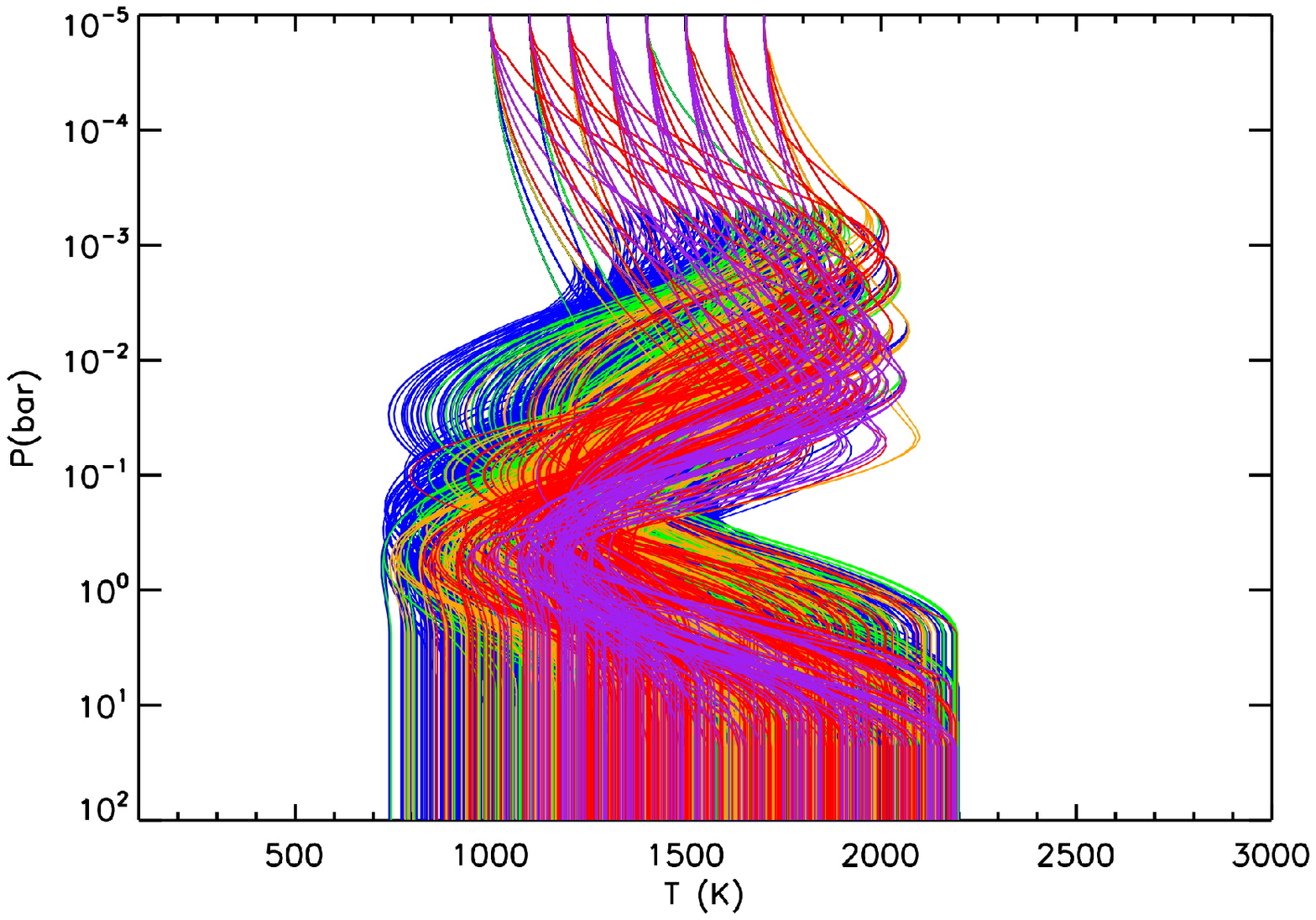}
\caption{Pressure-Temperature profiles explored by models for the
secondary eclipse spectrum of HD 209458b. All the profiles have an
inversion layer. Profiles without thermal inversions are ruled out by
the data. The profiles are color-coded by minimum $\xi^2$ (as described 
in Figure~\ref{fig:pt_hd189}), and the colors are described in 
Figure~\ref{fig:c_hd209}.}
\label{fig:pt_hd209}
\end{center}
\end{figure}

\section{HD 209458b Results: Secondary Eclipse}
\label{sec-hd209}

The dayside of HD~209458b has been of substantial interest, owing to
the indication of a thermal inversion in the atmosphere of this
planet.  Deming et al. (2005) reported the first detection of thermal
emission from HD~209458b in the 24 $\micron$ MIPS Channel of
\textit{Spitzer}.  Knutson et al. (2008) reported broadband
photometry of the planet-star flux ratio in the four \textit{Spitzer}
IRAC channels (3.5, 4.6, 5.8, 8 $\micron$).  Newer photometric
observations have been made by Deming in the 16 $\micron$ IRS Channel
and, an updated value is available for the 24 $\micron$ MIPS Channel
(Deming, private communication 2009). Richardson et al. (2003) observed an
upper-limit on the contrast ratio in the K-band (2.2 $\micron$) using
IRTF SpeX. Additionally, Richardson et al. (2007) and Swain et al. (2008a) presented a
low S/N \textit{Spitzer} IRS spectrum in the 7.5-13.2 $\micron$
range. And, after submitting our present work, we became aware of an independent
inference of H$_2$O, CH$_4$, and CO$_2$ in HD~209458b, based 
on {\it HST} NICMOS observations, and five-channel {\it Spitzer} 
photometry; although, a quantitative measure of fit was not reported 
(Swain et al. 2009b). Here, we report independent detections of H$_2$O, 
CH$_4$, CO$_2$, and CO based on quantitative constraints at the $\xi^2 < 1$ 
level of model fits to six-channel {\it Spitzer} photometry.

For HD~209458, we compute model fits to the 
six \textit{Spitzer} broadband photometric detections (at 3.6, 4.5, 5.8, 8, 
16 and 24 $\micron$), combining data from Knutson et al. (2008) and 
Deming (private communication, 2009). We do not use the IRS spectrum 
(Richardson et al. 2007; Swain et al. 2008a) because the S/N is too low to yield
signficant model constraints.  

We explore a grid of $6 \times 10^6$ models for this system.
Figure~\ref{fig:spec_hd209} shows sample model spectra and the data 
for the planet-star flux contrast in this system. The figure shows model
spectra at different levels of $\xi^2$. As shown in the figure, our best
fit spectra (a sample binned spectrum is shown in light blue) fit all
the available data to within the 1-$\sigma$ error bars. And, our
$\xi^2 = 2$ models match the observations to $ \sim 1.4 \sigma$ on
average.

\subsection{Molecular Abundances}

Our results indicate the presence of H$_2$O, CO, CH$_4$ and CO$_2$ in the
atmosphere of HD 209458b. As will be discussed in \S~\ref{sec-hd209-pt}, the 
allowed $P$-$T$ profiles show 
the existence of a deep thermal inversion in the atmosphere of HD 209458b.
Consequently, all the line features of the molecules are seen as emission 
features, rather than absorption features as in case of HD 189733b. The constraints 
on the concentrations of all the molecules are shown in Figure~\ref{fig:c_hd209}. 

The $\xi^2 = 2$ surface places an upper-limit on the mixing ratio of H$_2$O at
$10^{-4}$, with the $\xi^2 = 1$ surface allowing mixing ratios up to $10^{-5}$.  
The limits on H$_2$O are governed by the band features of the molecule in 
the 5.8 $\micron$ and 24 $\micron$ channels. While the upper limit on 
the mixing ratio of H$_2$O is well constrained, the constraint on the 
lower limit is rather weak, allowing values as low as $10^{-8}$ at $\xi^2 = 1$. 
The stringent upper limit on 
H$_2$O comes primarily from the low planet-star flux contrast observed in the 24 
$\micron$ channel. Similarly, the high flux contrast observed in the 5.8 $\micron$ 
channel should, in principle, yield a stringent lower limit. However, the 
weak lower limit is because of the large observational uncertainty on the
flux contrast in the 5.8 $\micron$ channel. The $1-\sigma$ error bar on the 
observed 5.8 $\micron$ flux contrast is the largest of uncertainties in all 
the channels, and 1.5 times larger than the error bar on the 24 $\micron$ flux contrast. 

The $\xi^2 = 2$ constraint on the mixing ratio of CH$_4$ is $10^{-8}$ - 0.04, based
on band features in the 3.6 $\micron$ and 8 $\micron$ channels.  Although, the features in
the 3.6 $\micron$ channel are degenerate with some features of H$_2$O in 
the same channel, the exclusive CH$_4$ features in the 8 $\micron$ channel break 
the degeneracy. The $\xi^2 = 2$ surface places an absolute lower-limit on CO at 
$4 \times 10^{-5}$, and requires a mixing ratio of CO$_2$ between 
$2 \times 10^{-9} - 7 \times 10^{-6}$. The need for CO$_2$ arises from the high 
contrast observed between the 4.5 $\micron$ and 
the 3.6 $\micron$ channels. The 4.5 $\micron$ channel could in principle be 
explained by CO alone. However, such a proposition would require large 
amounts of CO, close to mixing ratios of 1. CO$_2$ provides additional
absorption in the same 4.5 $\micron$ channel, with a relatively 
reasonable abundance. In addition, the amount of CO$_2$ is also constrained 
by the 16 $\micron$ channel --- too high of a CO$_2$ abundance would not
agree with the data because CO$_2$ has a strong absorption feature in the 
16 $\micron$ channel. Thus, the 4.5 $\micron$ and 16 $\micron$ channels 
together constrain the amounts of CO$_2$, thereby also constraining the 
CO feature in the 4.5 $\micron$ channel.  If the 16 $\micron$ data did not 
exist, there would have been a large degeneracy in the abundances 
of CO and CO$_2$. 

If we consider the $\xi^2 = 1$ surface, instead of the $\xi^2=2$
surface, our results place tighter constraints on all the
molecules. At this surface, the constraints on the mixing ratios are: 
$10^{-8} \leq$ H$_2$O  $\leq 10^{-5}$; $4 \times 10^{-8} \leq $ CH$_4$ $ \leq 0.03$; 
CO $\geq 4 \times 10^{-4}$; and, $4 \times 10^{-9} \leq $ CO$_2$ $ \leq 7 \times 10^{-8}$. 
However, in considering the $\xi^2 = 1$ surface, we might be over-interpreting the data.

We find definitive detections of CH$_4$, CO and CO$_2$. Even at the 
$\xi^2 = 4$ surface, the data requires non-zero mixing ratios for the 
carbon-bearing molecules. Furthermore, 
the results suggest a high C/O ratio (see Seager et al. 2005). Although 
the model fits give C/O ratio close to 1 or higher, our molecular abundance grid 
resolution is not high enough to put a precise lower limit to the C/O ratio. As seen from
Figure~\ref{fig:c_hd209}, the allowed models reach values as high as
 $\sim$ 60 for $\xi^2 = 1$, and $\sim$ 600 for $\xi^2 = 2$. As discussed
previously, and in \S~6.1, high molecular concentrations and high C/O
ratios could be localized effects overrepresented by the 1D averaged
retrieval.

\subsection{Temperature Structure}
\label{sec-hd209-pt}
The most interesting aspect about the atmosphere of HD 209458b is the
presence of a thermal inversion on the planet dayside, as previous
studies have pointed out (e.g., Burrows et al. 2008). While modeling
atmospheric spectra, one often encounters a degeneracy between
temperature structure and molecular compositions. However, by
exploring a large number of $P$-$T$ profiles we find that the day-side
observations for HD 209458b cannot be explained without a thermal
inversion, for any chemical composition, at the $xi^2 < 2$ level. On the 
other hand, our results place stringent constraints on the 1D averaged 
structure of the required thermal inversion.

Our results presented here explore $\sim$ 7000 $P$-$T$ profiles with thermal 
inversions. Figure~\ref{fig:pt_hd209} shows all the $P$-$T$ profiles with 
$\xi^2 \leq 5$. The ranges in the $P$-$T$ 
parameters explored are shown in Figure~\ref{fig:c_hd209}. In the presence of 
a thermal inversion, the spectra are dominated by the location and extent of the 
inversion layer, quantified by the parameters $P_2$, $T_2$, and $\alpha_2$.
We find the models with $\xi^2 \leq 2$ to have an inversion layer in the lower 
atmosphere, at pressures ($P_2$) above $\sim 0.03 $ bar, with the best-fit models 
preferring the deeper layers, i.e., higher $P_2$. We find that the $\xi^2 \leq 1 $ region extends to 
inversion layers as deep as $P_2 \sim 1$ bar. However, we limit $P_2$ to a value 
of 0.5 based on the physical consideration that, at pressure $\gtrsim$ 1 bar collisional 
opacities are likely to dominate over the molecular opacities in the optical which are 
expected to be the cause of thermal inversions. And, therefore, we do not expect
inversions at such high pressures. 

The allowed models (Figure~\ref{fig:c_hd209}) cover a wide range in
the base temperature ($T_3$) owing to the fact that the
spectrum is dominated by the maximum temperature differential ($T_1$ -
$T_2$) attained in the stratosphere, and is less sensitive to $T_3$ by
itself. The stratospheric temperatures of all the models with $\xi^2 \leq 2$
lie between 800 K and 2200 K. The temperature gradient in the inversion
layer is governed by the $\alpha_2$ parameter. As can be seen from
Figure~\ref{fig:c_hd209}, the data restricts the $\xi^2 \leq 2$ models to
have $\alpha_2$ between 0.05 and 0.25, with the $\xi^2 \leq 1$ models
preferring values between 0.05 and 0.15. These values of $\alpha_2$ translate 
into mean temperature gradients ($<dT/d \log P>$) in the inversion layer 
to be 230 K - 1480 K for $\xi^2 \leq 2$, and 400 K - 1480 K 
for $\xi^2 \leq 1$. Consequently, large thermal inversions spanning 
more than a few dex in pressure, and hence with 
less steeper thermal gradients, are ruled out by the data. This is 
also evident from Figure~\ref{fig:pt_hd209}. Finally, $\xi^2$ is 
not very sensitive to the parameters in Layer 1 of the $P$-$T$ 
profile, since the dominant spectral features in this planet arise 
from the inversion layer. 

\subsection{Albedo and Energy Redistribution}
Our results also show the constraints on the albedo and day-night
redistribution. Figure~\ref{fig:c_hd209} shows the $\xi^2$ surface 
in the space of $T_{\rm eff}$ vs. $\eta$, where, $\eta = (1-A_B)(1-f_r)$; 
 $A_B$ is the Bond albedo, and $f_r$ is the fraction of input stellar flux 
 redistributed to the night side. As described in \S~\ref{sec-hd189-fr}, 
 using $\eta$ one can estimate the constraint on $A_B$ given $f_r$, 
 or vice-versa. For HD~209458b, the $\xi^2 = 2$ surface in Figure~\ref{fig:c_hd209} 
constrains $\eta$ in the range 0.24 - 0.81, with the $\xi^2 = 1$ surface in the 
range 0.26 - 0.72. If we assume an $A_B$ of 0.24 (2-$\sigma$ upper-limit, scaled 
from the 1-$\sigma$ estimate of Rowe et al. 2008), the constraints on $f_r$ for the 
$\xi^2 = 2$ and $\xi^2 = 1$ surfaces are 0.0 - 0.68 and 0.05 - 0.66, respectively. 
This range of values allows for very efficient redistribution in the atmosphere 
of HD~209458b, although values of $f_r > 0.5$ seem physically unrealistic. 

\section{Discussion}
 \label{sec-discussion}

Our results in this work have been presented as goodness-of-fit 
contours in the parameter space of a 1D averaged atmosphere
model. In this section, we first discuss the interpretation of our results 
in the light of the fact that the atmosphere is inherently 3D, with 
dynamical effects that cannot be captured in a 1D model. 
We then present an assessment of our best-fit models vis-a-vis 
the goodness-of-fit contours. Finally, we discuss possible future 
extensions to our model. 

\subsection{Contribution Functions}

We have constrained the $P$-$T$ profile and abundances on the  
planetary day side. But which part of the atmosphere is really being
sampled by the observations? Here we will discuss the contribution to
the net emergent flux from the surface elements at the top of the atmosphere 
(see Knutson et al. 2009b for a discussion of radial contribution functions). We 
begin with the simple picture of radiative equilibrium, with no atmospheric 
circulation, in a homogeneous and isotropically radiating atmosphere. In this case, the
amount of flux from a given surface element is proportional to the 
amount of incident flux from the star, where the incident stellar flux
drops off as $\cos \theta$; $\theta$ is the angle away from the
sub-stellar point. In this picture, the planetary infrared flux is
greatest at the sub-stellar point, where the planet is being heated the
most. The area contribution to the net planet flux, however, increases 
as $\sin\theta$. The bigger annuli correspond to larger areas on the 
projected planetary disk as viewed by the
observer. The lower flux emitting regions (the outer regions of the disk,
with lower incident stellar intensity), therefore, have more area than 
the higher flux regions. There is a sweet spot at 45 degrees where the 
contribution to the total observed flux peaks, with a distribution of
$\sin(2\theta)$. In other words, the total flux is an average over a
wide region on the projected planetary disk.
 
A more realistic scenario is one not limited to radiative equilibrium;
indeed atmospheric circulation models show dramatic differences of 
emergent flux across the planetary disk caused by strong hydrodynamic
flows (see, e.g., Showman et al. 2009; Langton and Laughlin, 2008; Cho
et al. 2003; Dobbs-Dixon and Lin. 2008). Without knowing the atmospheric circulation 
and the resulting surface pattern, can we say anything about which part of the
day-side planetary disk our $P$-$T$ profiles and abundances
preferentially sample? Given a retrieved abundance for a molecule, we do
not know if this is from a localized concentration or from a global
abundance, or something in between. For instance, we have found
evidence for a high CO$_2$ abundance in HD 189733b (see also Swain et
al. 2009a). But is this just a local effect of transported
photochemical products? Or is the whole day-side evenly enriched in
CO$_2$?  If we are not measuring a true average, we have to be careful
about how we interpret the retrieved CO$_2$ abundance.

A thorough examination of the contribution of different parts of 
the planetary disk has to take into account the emergent surface flux profiles
that are more realistic than the assumption of homogeneity and 
radiative equilibrium. For example, 3-D models of hot Jupiter atmospheres 
by Showman et al. (2009) show a much slower than $\cos\theta$ drop 
off from the sub-stellar point caused by hydrodynamic flows. 
And at the planetary photosphere the ``sub-stellar hot spot" is blown 
downwind, to 30 degrees; a detail that would not be apparent in 
the assumption of an isotropic and homogenous atmosphere. Moreover, 
the {\it Spitzer} 8 micron measurements by Knutson et al. (2007) 
show a hotspot at lower separation from the sub-stellar point (and also 
show a cold spot on the same hemisphere). For instance, 
if CO$_2$ were predominantly formed in localized regions and blown 
downwind, it could land at geometrically favored regions in terms of 
the 1D averaged spectrum. Alternatively, vertically localized 
concentrations of CO$_2$ at favorable pressure levels could give 
rise to strong spectral signatures, similar to those formed by 
having the same concentration of CO$_2$ over the entire 
atmosphere. In these cases, a high value of CO$_2$ may not be 
representative of the entire day side.  
  
\subsection{Best Fit Models vs. Error Surface}
\label{sec-discuss-chisqr}
In the current work we have described most of our results at the
$\xi^2=2$ level. Our formulation for $\xi^2$, as described in 
\S~\ref{sec-xisqr}, would be equivalent to the conventional definition of 
reduced $\chi^2$, if $N_{obs}$ were to be replaced by the number of degrees of 
freedom. Normally, one aims for fits at the reduced $\chi^2=1$ level (we
have also provided results for fits at the level of $\xi^2=1$).  A $\chi^2$,
however, is devoid of meaning when the number of degrees of freedom is
less than zero---in that we cannot relate $\chi^2$ to a confidence level.
In the absence of confidence levels, we are considering $\xi^2$ as
a weighted mean squared error. In this framework, a $\xi^2=1$ means that,
on average, the model deviates from the mean values of the data by
1-$\sigma$, and $\xi^2=2$ means that the deviation is $\sqrt{2}$ = 1.4
$\sigma$. In this work, the number of data points in the {\it Spitzer}
broadband photometry is less than the number of free model
parameters, meaning less than zero degrees of freedom. In both the
{\it Spitzer} IRS and the {\it HST}/NICMOS data sets, there are, in principle, more data 
points than the number of free model parameters. But both these data sets
sample a narrow wavelength range, with a limited number of absorption features. 
In addition, the degeneracies in our model parameter space are poorly 
understood, if at all.  Without a robust parameter-fitting algorithm, we cannot 
assign confidence levels even with the {\it HST} and IRS data sets. Which error 
surface should be used to encompass the best fit models, given that we have no 
direct correspondence to confidence levels? There is no conclusive answer and, 
ideally, one might like to consider only those models which fit the data to within 
$\xi^2=1$, or even $\xi^2=0$ indicating perfect fits. However, we have used 
a conservative choice of the $\xi^2=2$ surface for interpretation of our results, 
allowing for models fitting the observations to within $\sim 1.4 \sigma$ of each 
data point, on an average.

One approach to accurately consider the uncertainty on the data is as
follows. One could take tens of thousands of realizations of each data
set, uniformly sampled over the error distribution. Fitting each realized 
data set would mean running millions of models for each case, leading 
up to billions of models per data set per planet. Such an approach is 
computationally prohibitive at the present time.

\subsection{Suggested Variability}
\label{sec-variability}

There is evidence that the dayside atmosphere of HD~189733b is variable. 
One highly relevant example to our interpretation is the 8 $\micron$ {\it Spitzer}
IRAC measurement of HD~189733b reported by Knutson et al. (2007) 
versus that reported by Charbonneau et al. (2008) in the same channel at a 
different time. The flux ratios differ by $\sim$ 2.3 $\sigma$. Also significant is 
the disagreement in the planet-star contrast levels between the HD~189733b 
IRS spectrum and the IRAC broadband photometry at 8 $\micron$, but not at 
5.8 $\micron$ (Grillmair et al. 2008).

Atmospheric variability is also suggested by the very strong CO$_2$ feature 
in the HD~189733 HST/NICMOS data set and the lack of a noticeably strong 
absorption feature in the 4.5~$\micron$ and 16~$\micron$ Spitzer channels. 
It is well known that the CO$_2$ molecule has an extremely strong absorption 
feature at 15 $\micron$; yet there is no indication of strong CO$_2$ absorption 
observed in the 16 $\micron$ channel. Whether this data variability is due to 
intrinsic variability of the planet atmosphere or instead is due to the different 
data reduction techniques is under study (Desert et al. 2009; Beaulieu et al. 2009). 

Turning to our model results, we have been unable to find fits to all three 
data sets of HD~189733b at the $\xi^2 \le 2$ level, in the large range of 
parameter space we have explored. There may be befitting $P$-$T$ profiles 
and compositions that our grid did not cover, but we think this is unlikely, 
mainly because the data sets show noticeable differences even without 
running the models, e.g. the observed features of CO$_2$. We plan to 
implement a more exhaustive exploration of the $P$-$T$ parameter space 
by extending our work with a new optimization technique in the future. 

\subsection{Model Extensions}
\label{sec-model_ext}

Our approach allowed us to run a large ensemble of models in the 
parameter space. Nevertheless, future calculations are needed to 
estimate the covariances between the model parameters. Several 
model parameters are  degenerate; for example, molecular abundances 
of species that share absorption features in the same photometric 
channel (e.g., CO and CO$_2$). Also well-known is the degeneracy 
between the $P$-$T$ profile and the molecular abundances. Understanding 
the covariances between the parameters might help break the 
degeneracies in the future. 

We chose the $\sim 10^4$ $P$-$T$ profiles based on physical arguments and
preliminary fits to the data. Even though having a grid of $\sim 10^4$ $P$-$T$ profiles 
in the six-parameter $P$-$T$ space, and $\sim 10^7$ models overall, means that our 
grid is coarse, we believe it covers most physically plausible profiles. Nonetheless, a 
more complete exploration requires a more sophisticated approach, and may find 
$P$-$T$ profiles and the corresponding abundance constraints that may be slightly 
different from those reported here. 

Our model does not include clouds or scattering. Although cloud-less models
are justified for hot Jupiters (see \S~\ref{sec-molecules}), clouds are needed 
for extending our model to cooler planetary atmospheres, such as those of 
habitable zone super Earths and cooler giant planets. Our model also does 
not include photochemistry, but instead varies over different molecular 
abundances. Once a set of  best-fit models are found by our approach, the 
molecular abundances can be used to guide photochemical models; conversely, 
photochemical models can be used to check the plausibility of the best-fit 
molecular abundances. And, although we have used the latest available 
high-temperature opacities, further improvements are sought, 
particularly for CH$_4$ and CO$_2$. 

Our model presented in this work is suitable only for interpreting the 
infrared spectra of planetary atmospheres; it was motivated by existing observations. 
We do not include the sources of opacity and scattering that are required to 
model the visible part of the spectra. However, our model can be extended 
to visible wavelengths, and we aim to discuss it in a future work. Constraints 
placed by visible observations will help to further narrow down the model 
parameter space already constrained by IR observations.

We have not included the intrinsic flux from the planet interior, because, for 
hot Jupiters stellar irradiation is the dominant energy source. 
However, extensions of our model to brown dwarf atmospheres or young, or 
massive, Jupiters will require inclusion of the interior energy. And, although 
irradiation is naturally included in the form of the parametric $P$-$T$ profile, 
our model gives no information about the nature of the absorber. Nevertheless, 
given the best-fit $P$-$T$ profile from our approach, one could quantify the 
range of parameters of a potential absorber.

\subsection{Temperature Retrieval as a Starting Point for Forward Models} 
\label{sec-retrieval}
We have presented a radical departure from the standard exoplanet
atmosphere models. Our temperature and abundance retrieval method does
not use the assumption of radiative + convective equilibrium in each 
layer to determine the $P$-$T$ profile. By searching over millions 
of models, we are extracting the $P$-$T$ profile from the data. 

We have not necessarily suggested to replace so-called
``self-consistent forward models", those models that include: radiative
transfer, hydrostatic equilibrium, radiative + convective
equilibrium, day-night energy redistribution, non-equilibrium chemistry, 
and cloud formation. Instead of a stand-alone model, our method can 
be used as a starting point for narrowing the potential parameter space 
to run forward models in. We anticipate that this temperature retrieval 
method will enable modelers to run enough  forward models to find 
quantitative constraints from the data, instead of running only a 
few representative models.
 
Atmospheric temperature and abundance retrieval is not new. Solar
system planet atmosphere studies use temperature retrieval methods,
but in the present context there is one major difference. Exquisite
data for solar system planet atmospheres means that a fiducial $P$-$T$ 
profile can be derived. The temperature retrieval process for 
a solar system planet atmosphere therefore involves perturbing the
fiducial profile. For exoplanets, there is no starting point to derive
a fiducial model because of the limited data. Building on methods 
used for solar system planets, Swain et al. (2008b \& 2009) have used 
a temperature retrieval method using published $P$-$T$ profiles 
from other models, and perturbing the profiles to find fits to data 
(also see Sing et al. 2008). In this work, we report a new approach for atmospheric retrieval of 
exoplanets which allows us to explore a wide region of the parameter 
space, and to possibly motivate a new generation of retrieval methods 
tailored to exoplanet atmospheres.

\section{Summary and Conclusions}
 \label{sec-summary}

We have presented a solution to the retrieval problem in exoplanet
atmospheres: given a set of observations, what are the constraints on
the atmospheric properties?  Our method consists of a new model 
involving line-by-line radiative transfer with a parametric pressure-temperaure ($P$-$T$) 
profile, along with physical constraints of hydrostatic equilibrium and 
global energy balance, and allowing for non-equilibrium molecular 
compositions. This approach enabled us to run millions of models of 
atmospheric spectra over a grid in the parameter space, and to calculate 
goodness-of-fit contours for a given data set. Thus, given a set of 
observations and a desired measure of fit, our method places 
constraints on the $P$-$T$ profile and the molecular abundances in 
the atmosphere. 

We used our method to place constraints on the atmospheric properties 
of two hot Jupiters with the best available data: HD~189733b 
and HD~209458b. We computed $\sim 10^7$ models for each system.
Our constraints on the $P$-$T$ profiles confirm the presence of a thermal 
inversion on the dayside of HD~209458b, and the absence of one in
HD~189733b, as has been previously reported in the literature. 
And, the constraints on molecular abundances 
confirm the presence of H$_2$O, CH$_4$, CO and CO$_2$ in 
HD~189733b, as has been found by several other studies in the past. 

We presented some new findings. We report independent detections 
of H$_2$O, CH$_4$, CO$_2$, and CO, from six-channel {\it Spitzer} 
photometry of HD~209458b, at the $\xi^2 < 1$ level (also see Swain et al. 2009b).
At the $\xi^2 = 2$ surface, we find the lower-limits
on the mixing ratios of CH$_4$, CO and CO$_2$ to be $10^{-8}$, 
$4 \times 10^{-5}$ and $2 \times 10^{-9}$, respectively. In
addition, we find an upper-limit on H$_2$O mixing ratio of $10^{-4}$. Secondly, our
results using the high S/N IRS spectrum of HD 189733b 
indicate that the best fit models are 
consistent with efficient day-night circulation on this planet,
consistent with the findings of Knutson et al. (2007 \& 2009b); 
although, we also find that the broad band photometry data 
requires much less efficient day-night circulation, as has been 
reported by several studies in the past. The different 
constraints placed by the different data sets arise from the 
variability in the data. 

A few key observations of HD~189733b vary between different data 
sets at similar wavelengths. Moreover, a noticeably strong CO$_2$ 
absorption in one data set is significantly weaker in another. We must, 
therefore, acknowledge the strong possibility that the atmosphere is variable, 
both in its energy redistribution state and in the chemical abundances.

Our new temperature and abundance retrieval --- together with the limited
but remarkable data sets --- charts a new course for exoplanet atmosphere
interpretation. Whether our approach is used as a stand alone model, 
or as a starting point for forward models, the field of exoplanet atmospheres 
is ready for a technique to place detailed atmospheric constraints governed by the data. 
The ideal data set is one with a wide wavelength range that covers more 
than one spectral feature of different molecules. The {\it James Webb Space Telescope}, 
scheduled for launch in 2013, is ideal for observations of transiting exoplanet atmospheres
and should usher a new era for quantitative analyses of exoplanet atmospheres.

\acknowledgements{We thank Drake Deming, Saul Rappaport, Josh Winn, 
Tamer Elkholy, Josh Carter and Adam Burgasser for very helpful discussions. 
We thank Richard Freedman for providing molecular line lists, especially
with helpful information on CO$_2$ opacities. We thank Larry Rothman for access to the HITEMP database. We thank Scott Hughes and Paul Hsi for facilitating access to the 
computer cluster at the MIT Kavli Institute. Support for this work was provided by 
NASA through an award issued by JPL/Caltech.}

\end{document}